\pdfoutput=1
\documentclass[aps,prd,amsmath,floats,floatfix, twocolumn,
superscriptaddress,nofootinbib,showpacs,longbibliography]{revtex4-1}
\usepackage[T1]{fontenc}
\usepackage[utf8]{inputenc}
\usepackage{lmodern}
\usepackage{verbatim}
\usepackage[dvipsnames, usenames]{xcolor}
\definecolor{linkcolor}{rgb}{0.0,0.3,0.5}
\usepackage[hypertexnames=false, unicode, colorlinks=true, linkcolor=linkcolor,
citecolor=linkcolor, filecolor=linkcolor,urlcolor=linkcolor,
pdfusetitle]{hyperref}
\usepackage[all]{hypcap}
\usepackage{graphicx}
\usepackage{xspace}
\usepackage{amssymb}
\usepackage{microtype}

\usepackage{url}

\usepackage[english]{babel}
\usepackage{blindtext}

\usepackage[normalem]{ulem} 
\usepackage{bm} 
\usepackage{mathrsfs}

\usepackage[caption=false]{subfig}

\DeclareMathAlphabet{\mathpzc}{OT1}{pzc}{m}{it}

\newcommand{\gu}[1]{\textcolor{orange}{#1}}

\usepackage[nomessages]{fp}

\def\di{\partial}
\def\mF{\mathcal{F}}

\def\pphi{p_\varphi}
\def\pphidot{\dot{p}_\varphi}

\begin{document}
\title{Phenomenology and origin of late-time tails in eccentric binary black hole mergers}
\newcommand{\KITP}{\affiliation{Kavli Institute for Theoretical Physics, University of California Santa Barbara, Kohn Hall, Lagoon Rd, Santa Barbara, CA 93106}}
\newcommand{\UMassDMath}{\affiliation{Department of Mathematics,
		University of Massachusetts, Dartmouth, MA 02747, USA}}
\newcommand{\UMassDPhy}{\affiliation{Department of Physics,
		University of Massachusetts, Dartmouth, MA 02747, USA}}
\newcommand{\CSCVRUMass}{\affiliation{Center for Scientific Computing and Data Science Research, University of Massachusetts, Dartmouth, MA 02747, USA}}
\newcommand{\URI}{\affiliation{Department of Physics and Center for Computational Research, 
    University of Rhode Island, Kingston, RI 02881, USA}}    
\newcommand{\AEI}{\affiliation{Max Planck Institute for Gravitational Physics (Albert Einstein Institute), Am M¨uhlenberg 1, Potsdam, 14476, Germany}}
\newcommand{\NBIA}{\affiliation{Niels Bohr International Academy, Niels Bohr Institute, Blegdamsvej 17, 2100 Copenhagen, Denmark}}
\author{Tousif Islam}
\email{tislam@kitp.ucsb.edu}
\KITP
\UMassDPhy
\UMassDMath
\CSCVRUMass

\author{Guglielmo Faggioli}
\AEI 

\author{Gaurav Khanna}
\URI
\UMassDPhy
\CSCVRUMass

\author{Scott E. Field}
\UMassDMath
\CSCVRUMass

\author{Maarten van de Meent}
\NBIA
\AEI

\author{Alessandra Buonanno}
\AEI

\hypersetup{pdfauthor={Islam et al.}}

\date{\today}

\begin{abstract}
We investigate the late-time tail behavior in gravitational waves from merging eccentric binary black holes (BBH) using black hole perturbation theory. For simplicity, we focus only on the dominant quadrupolar mode of the radiation. We demonstrate that such tails become more prominent as eccentricity increases. Exploring the phenomenology of the tails in both spinning and non-spinning eccentric binaries, with the spin magnitude varying from $\chi=-0.6$ to $\chi=+0.6$ and eccentricity as high as $e=0.98$, we find that these tails can be well approximated by a slowly decaying power law. We study the power law for varying systems and find that the power law exponent lies close to the theoretically expected value $-4$. Finally, using both plunge geodesic and radiation-reaction-driven orbits, we perform a series of numerical experiments to understand the origin of the tails in BBH simulations. Our results suggest that the late-time tails are strongly excited in eccentric BBH systems when the smaller black hole is in the neighborhood of the apocenter, as opposed to any structure in the strong field of the larger black hole. 
Our analysis framework is publicly available through the \texttt{gwtails} Python package.
\end{abstract}

\maketitle

\section{Introduction}

Understanding binary black hole (BBH) coalescence is a key to gravitational wave (GW) astronomy. The coalescence of binary black hole (BBHs) is typically characterized by different distinctive regimes starting from the inspiral and culminating in the ringdown stage. The ringdown of a BBH merger is primarily dominated by quasi-normal modes (QNM) ringing in the early times and has been extensively studied using both linear black hole perturbation theory (BHPT) and fully nonlinear numerical relativity (NR) simulations~\cite{London:2018nxs,Berti:2014fga,Berti:2005ys,Berti:2005ys,Baibhav:2023clw,London:2018gaq,Redondo-Yuste:2023seq}. These studies not only have provided a phenomenology of the QNMs but also has offered analytical templates.
However, the late-time behavior of a ringdown signal has primarily been studied within the BHPT framework~\cite{Price:1971fb, Price:1972pw}.  These studies have indicated the existence of a slowly-decaying power-law-controlled tail -- so-called ``Price tail'' in late-time evolution. Subsequently, considerable efforts have been invested in understanding the late-time power-law tail behaviors in Schwarzschild spacetime as well as in the Kerr case within the BHPT framework~\cite{Gundlach:1993tp, Gundlach:1993tn, Okuzumi:2008ej,Burko:1997tb,Barack:1998bw,Bernuzzi:2008rq,Burko:2013bra, Zenginoglu:2012us, Burko:2010zj, Burko:2004jn, Burko:2007ju, Krivan:1999wh, Poisson:2002jz, Burko:2002bt, Barack:1999ma,Racz:2011qu, Harms:2013ib}. In particular, the asymptomatic behaviors of these tails are studied in detail in Ref.~\cite{Zenginoglu:2009ey}.

Only recently, Ref.~\cite{Albanesi:2023bgi,DeAmicis:2024not} employed perturbative techniques within the Regge-Wheeler-Zerilli (RWZ) framework~\cite{PhysRev1081063,PhysRevLett24737,PhysRevD71104003,Nagar:2005ea} to simulate non-spinning BBH mergers with eccentricities. They have observed that the late-time tail behavior is larger and occurs much earlier than in their quasi-circular counterparts. Soon after, Ref.~\cite{Carullo:2023tff} has noticed hints of similar eccentricity-induced tails in comparable-mass non-spinning eccentric binaries using publicly available RIT NR data~\cite{Healy:2022wdn}. The relatively shorter length of the NR data, however, makes it difficult for the authors to probe the tail behavior in detail.

In this paper, we aim to provide a detailed phenomenology of the eccentricity-induced tails for both non-spinning and spinning binaries using a BHPT approach based on the Teukolsky equation~\cite{Sundararajan:2007jg,Sundararajan:2008zm,Sundararajan:2010sr,Zenginoglu:2011zz,WENO,Taracchini:2013wfa,Taracchini:2014zpa,Barausse:2011kb,Nagar:2006xv}. It is important to note that BHPT simulations are particularly suitable for this scenario. Firstly, we are probing the relaxation of the black hole created at the end of the merger, and thus, this is within the regime of validity of the BHPT framework. 
While linear BHPT framework cannot probe higher-order effects in ringdown~\cite{Mitman:2022qdl,Cheung:2022rbm}, these effects are expected to be small for sufficiently asymmetric binaries and can, therefore, be neglected.

In this paper, we simulate a set of highly eccentric BBH mergers with eccentricity (at the last stable orbit) ranging from $e=0.8$ and $e=0.98$. We also vary the spin of the larger black hole to study the impact of spin on the tail behavior. Furthermore, we develop and apply a framework to model the tail behaviors of the binary and compare our results within the existing literature. We make our analysis framework publicly available at \href{https://github.com/tousifislam/gwtails}{https://github.com/tousifislam/gwtails} for the ease of reproducibility.

The paper is organized as follows. Section~\ref{sec:framework} provides a detailed overview of our analysis framework. We describe our analytical template for the ringdown amplitude in Section~\ref{sec:model} and outline the method for extracting model parameters from the ringdown data in Section~\ref{sec:fitting_method}. We then look into the phenomenology of ringdown amplitudes in eccentric non-spinning binaries in Section~\ref{sec:nospin} and eccentric spinning binaries in Section~\ref{sec:spin}. Subsequently, we extract tail parameters and discuss their dependence on the eccentricity and spin values of the binary. We then perform a series of numerical experiments to understand the source of late-time tails in Section~\ref{sec:tail_reason}. Finally, in Section~\ref{sec:NR}, we examine NR data from both the SXS collaboration~\cite{Boyle:2019kee} and RIT catalog~\cite{Healy:2022wdn} to search for evidence of tails. Appendix~\ref{app:model} provides some intuition behind tail excitation and generation in the context of Schwarzschild spacetime using the RWZ equations.

\section{Analysis framework}
\label{sec:framework}
In this section, we first present an executive summary of the BHPT framework used in this paper. We then describe the analytical model we employ to describe the ringdown data. Subsequently, we provide a brief outline of the iterative framework used in fitting the data.

\subsection{Notation}
We adopt natural units $G=c=1$ and work in the center of mass frame of the binary. The mass of the larger (smaller) black hole is denoted by $m_1$ ($m_2$). We define the total mass of the system as $M:=m_1+m_2$; unless otherwise specified, we use $M=1$ throughout., the reduced mass as $\mu:=m_1m_2/M$ and the mass ratio as $q:=m_1/m_2$. For all of our BBH simulations, we use $\mu=10^{-3}M$. The dimensionless spin parameter of the Kerr black hole is defined as $\chi = J/M^2$ where $J$ is the total angular momentum. Additionally, we denote the retarded time as $t$. 

\subsection{BBH simulation using black hole perturbation theory}
\label{sec:BHPT_simulation}
We simulate BBH mergers within the BHPT framework using a time-domain Teukolsky solver. The smaller black hole is modeled as a point-particle, with no internal structure, moving in the spacetime of the larger Kerr black hole. Details of the framework are provided in Refs.~\cite{Sundararajan:2007jg,Sundararajan:2008zm,Sundararajan:2010sr,Zenginoglu:2011zz, WENO}. The framework first computes the trajectory taken by the point-particle using a test-mass effective-one-body (EOB) model~\cite{Faggioli} and then we use that trajectory to compute the gravitational wave emission. We start the simulations close to plunge (typically about 2 orbits before plunge) and let it evolve for a long time after the merger so that we can probe the tail effects effectively. 

\subsubsection{Trajectory}
\label{sec:trajectory}
We describe the dynamics of the point-particle orbiting the Kerr black hole using EOB formalism~\cite{Buonanno:1998gg, Buonanno:2000ef}. The trajectory of the point-particle is given by a set of four dynamical variables $\{ R, \varphi, P_R, P_{\varphi} \}$. Here, $R$ is the radial separation between the two black-holes, $\varphi$ is the orbital phase, $P_R$ is the radial momentum whereas $P_{\varphi}$ denotes the angular momentum. Subsequently, we define a set of dimensionless variables $\{r, \varphi, p_r, p_{\varphi} \}$ as:
\begin{gather}
  r=\frac{R}{M}, \quad p_r = \frac{P_r}{\mu}, \quad 
  p_{\varphi}=\frac{P_\varphi}{M\mu},
\end{gather}
and write the $\mu$-normalized Kerr Hamiltonian: 
\begin{equation} \label{Eq: Kerr_Hamiltonian}
H = \Lambda^{-1} \left(2 \chi p_{\varphi}+\sqrt{\Delta  p_{\varphi}^2 r^2 + \Delta^2 \Lambda  \frac{p_r^2}{r}+\Delta  \Lambda  r} \right) \ , 
\end{equation}
with dimensionless quantities $\Lambda$ and $\Delta$ being 
\begin{subequations}
\begin{align}
\Lambda & = r^3 + 2 \chi^2 + \chi^2 r \ , \\
\Delta & = r^2 - 2 r + \chi^2 \ .
\end{align}
\end{subequations}
We substitute the radial momentum $p_r$ with $p_{r_{*}}$, the momentum conjugate to the tortoise radial coordinate $r_{*}$. 
The tortoise coordinate is related to the Boyer-Lindquist $r$ by:
\begin{subequations}
\begin{align}
& dr_{*} = \frac{r^2 + \chi^2}{\Delta} dr = \frac{1}{\xi(r)} dr \ , \\
& p_{r_{*}} = \xi(r) p_r \ .
\end{align}
\end{subequations}
This is a general practice which is done to improve the numerical stability of the dynamics evolution, since 
$p_r$ diverges at the horizon while $p_{r_{*}}$ does not.
The evolution of the point-particle dynamics is provided by the Hamiltonian equations of motions: 
\begin{subequations}
\begin{align}
  & \dot{r} = \xi \frac{\di H}{\di p_{r_{*}}}(r, p_{r_{*}}, p_{\varphi}) \ , \label{Ham_EOM_1} \\
  & \dot{\varphi} = \frac{\di H}{\di \pphi} (r, p_{r_{*}}, p_{\varphi}) \ , \label{Ham_EOM_2} \\
  & \dot{p}_r = -  \xi \frac{\di H}{\di r}(r, p_{r_{*}}, p_{\varphi}) + \mF_r \ , \label{Ham_EOM_3} \\
  & \pphidot = - \frac{\di H}{\di \varphi}(r, p_{r_{*}}, p_{\varphi}) + \mF_{\varphi} \ . \label{Ham_EOM_4}
\end{align}
\end{subequations} 
where $\mF=(\mF_r, \mF_{\varphi})$ corresponds to the $\mu$-normalized radiation-reaction (RR) force connected to the emission of GWs for generic equatorial orbits. 
We obtain $\mF$ using a multiplicative resummation that contains eccentric corrections up to 3PN~\cite{Faggioli}.

We characterize the planar trajectories through the parameters $\{p, e, \chi \}$, which correspond to the semilatus rectum, the eccentricity and the spin of the Kerr BH.
We adopt the Keplerian parameterization where $p$ and $e$ are defined as:
\begin{equation}
p = \frac{2r_{\text{a}}r_{\text{p}}}{r_{\text{a}} + r_{\text{p}}} \quad , \quad e = \frac{r_{\text{p}} - r_{\text{a}}}{r_{\text{p}} + r_{\text{a}}} \,,
\end{equation}
with $r_{\text{a}}$ and $r_{\text{p}}$ being the radial separation at the apocenter and at the pericenter respectively. For the trajectories evolved with RR force, the eccentricity values are provided at the last stable orbit (LSO) crossing which occurs when the energy of the system equals the maximum of the radial potential $H(r, p_{r_*}  = 0, p_{\varphi})$.

\begin{figure}
\includegraphics[width=\columnwidth]{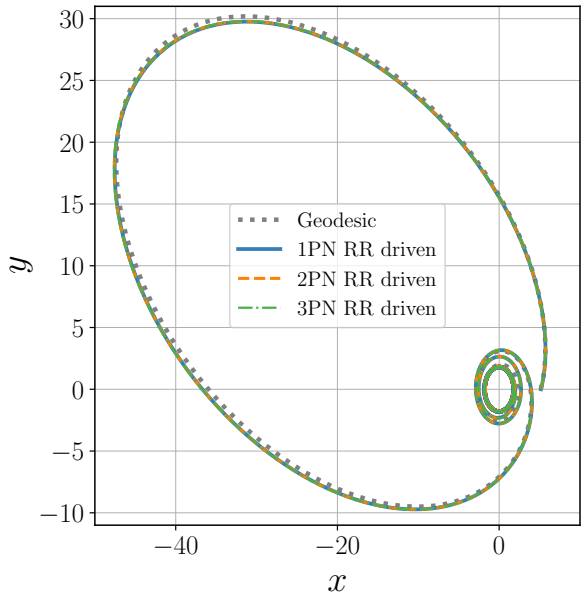}
\caption{We show the radiation-reaction driven trajectories and geodescic orbits of a plunging point-particle in a binary system with spin $\chi=0.6$ and eccentricity $e=0.9$. These orbits start with an initial energy of $E = 0.9823348$ and angular momentum $p_{\varphi}=3.1763217$. Unless stated otherwise, the 3PN RR driven inspiral orbit model is the default model used throughout this paper.
}
\label{fig:trajs_plot}
\end{figure}

In Figure~\ref{fig:trajs_plot}, we present the trajectories of a plunging point-particle in a binary system with spin $\chi=0.6$ and eccentricity $e=0.9$ for demonstration. We compute these trajectories using the radiation-reaction force of Ref.~\cite{Faggioli} at 1PN, 2PN, and 3PN order in the eccentric part. These trajectories are shown in the $xy$ plane where $x=r \cos \varphi$ and $y=r\sin \varphi$. For comparison, we also show the corresponding geodesic orbit. These simulations start with an initial energy of $E=0.9823348$ and angular momentum $p_{\varphi}=3.1763217$.

\subsubsection{Waveform generation}
\label{sec:wave_gen}
Once the trajectory of the perturbing compact body is fully specified as described above, 
we solve the inhomogeneous Teukolsky equation in the time-domain while feeding the trajectory 
information into the particle source-term of the equation \cite{Sundararajan:2007jg,Sundararajan:2008zm,Sundararajan:2010sr,Zenginoglu:2011zz,WENO}.
This involves a multi-step process: (i) rewriting the Teukolsky equation using compactified hyperboloidal 
coordinates that allow for the extraction of the gravitational waveform directly at null infinity while 
also solving the ``outer boundary problem'' of the finite computational domain; (ii) transforming the 
equation into a set of (2+1) dimensional PDEs by using the axisymmetry of the background Kerr space-time, 
and separating the dependence on azimuthal coordinate; (iii) recasting these equations into a first-order, 
hyperbolic PDE system; and lastly (iv) implementing a high-order WENO (3,5) finite-difference scheme with 
Shu-Osher (3,3) explicit time-stepping~\cite{WENO}. 

Once the Teukolsky solution is extracted at null infinity, it is straightforward to compute the complex strain $h$ by performing a double time-integral of the Weyl curvature scalar $\psi_4$. 

\begin{figure}
\includegraphics[width=\columnwidth]{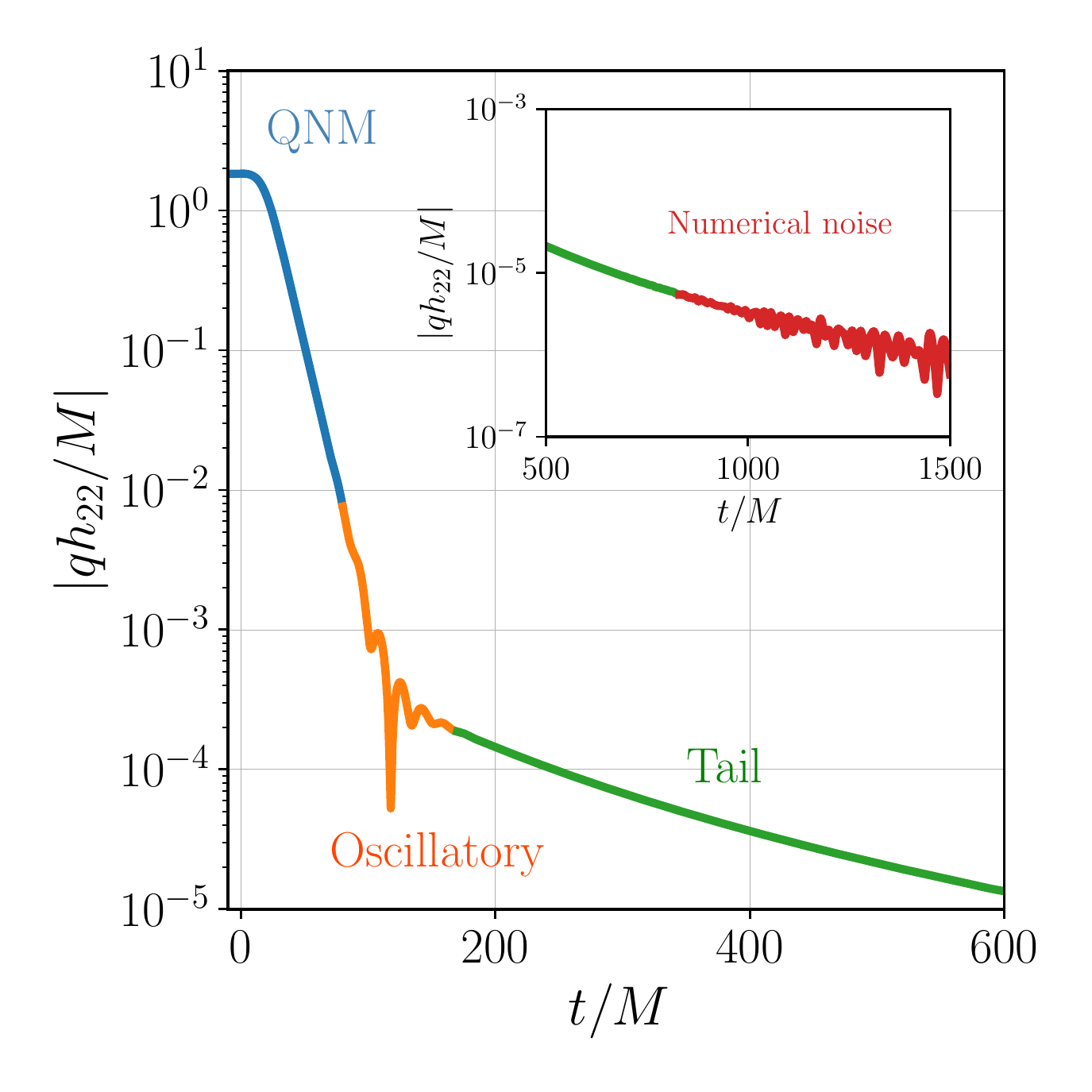}
\caption{We show the post-merger amplitude, exhibiting four distinct regimes, for a non-spinning binary with an eccentricity of $e=0.98$. These four regimes are (i) initial fast decaying QNM regime (blue), (ii) intermediate short oscillatory regime (orange), (iii) late-time regime with slowly varying tails (green), and (iv) a regime dominated by noise in numerical simulations (red, inset). More details are in Section~\ref{sec:fitting_method}.}
\label{fig:features}
\end{figure}

\subsection{Analytical model for the post-merger signal}
\label{sec:model}
Gravitational radiation (waveform) from a BBH merger is decomposed as a superposition of $-2$ spin-weighted spherical harmonic modes with indices $(\ell,m$):
\begin{align}
h(t,\theta,\phi;\boldsymbol{\lambda}) &= \sum_{\ell=2}^\infty \sum_{m=-\ell}^{\ell} h_{\ell m}(t;\boldsymbol\lambda) \; _{-2}Y_{\ell m}(\theta,\phi)\,,
\label{hmodes}
\end{align}
where $\boldsymbol{\lambda}$ is the set of intrinsic parameters (such as the masses and spins of the binary) describing the binary, and ($\theta$,$\phi$) are angles describing the orientation of the binary with respect to the observer.
Each spherical harmonic mode $h_{\ell m}(t)$ is a complex time series and is further decomposed into a real amplitude $A_{\ell m}(t)$ and phase $\phi_{\ell m}(t)$, as
\begin{equation}
h_{\ell m}(t) = A_{\ell m}(t) e^{i \phi_{\ell m}(t)} \,.
\label{eq:amp_phase}
\end{equation}
We choose the time axis in such a way that $t=0$ denotes the maximum amplitude of the $(2,2)$ spherical harmonic mode. 

While spherical harmonic modes are commonly used to model radiation from the inspiral to the ringdown, it is the ringdown waveform that offers richer phenomenology. Primarily, spheroidal harmonics provide a better description of the signal than spherical harmonics. Nevertheless, each spherical harmonic mode in the ringdown can be decomposed into a set of spheroidal harmonic modes (or in other words, quasi-normal modes (QNMs)), typically modelled by a superposition of damped sinusoidal. Additionally, ringdown signals also exhibit tail behaviors which can be modelled as a power-law decay. Each spherical harmonic mode can then be written as:
\begin{equation}
    h^{\rm ringdown}_{\ell  m} (t) = h^{\rm QNM}_{\ell  m} (t) + h^{\rm tail}_{\ell  m} (t).
\end{equation}
The QNM part of the ringdown $h^{\rm QNM}_{\ell  m}$ is given as the sum of all QNMs contribution to that $\ell m$ mode,
\begin{align}
     h^{\rm QNM}_{\ell  m} (t) & = 
     \sum_{\mathfrak{l}=2}^{\infty} \sum_{n=0}^{\infty} \mathcal{A}_{\mathfrak{l} m n} \, e^{-\frac{t}{\tau_{\mathfrak{l} m n}} 
     -i\omega_{\mathfrak{l} m n} t
     - i\phi_{\mathfrak{l} m n}  } 
     \nonumber \\
     &\quad+\sum_{\mathfrak{l}=2}^{\infty} \sum_{n=0}^{\infty} \mathcal{A}'_{\mathfrak{l} m n} \, e^{
     - \frac{t}{\tau'_{\mathfrak{l} m n}} 
     -i \omega'_{\mathfrak{l} m n} t
     -i \phi'_{\mathfrak{l} m n}  },
\end{align}
where $\omega_{\mathfrak{l} m n}$ and $\tau_{\mathfrak{l} m n}$ denote the charcteristic frequency and damping time of the $(\mathfrak{l} m n)$ (spheroidal) QNMs, and $\mathcal{A}_{\mathfrak{l} m n}$ and  $\phi_{\mathfrak{l} m n}$ are its amplitude and phase. The primes denote their ``mirror modes'', and the parameter $n$ denotes the overtones.
Typically, $n=0$ is known as the fundamental mode and carries most of the radiation.

The $h^{\rm tail}_{\ell  m}$ represents the ``tail''  contribution of the ringdown generated by the branch cut of the Green's function  along zero frequency axis. At sufficiently late times it is expected to behave as 
\begin{equation}\label{eq:htail}
    h^{\rm tail}_{\ell  m} (t) = \mathcal{A}_{\rm tail, \ell m}(t+c_{\rm tail, \ell m})^{p_{\rm tail, \ell m}} e^{i \phi_{\rm tail,\ell m}},
\end{equation}
with $p_{\rm tail, \ell m}=-(\ell+2)$~\cite{Barack:1999st,Hod:1999ry}. This means that, for the quadrupolar mode we are studying, we should have $p_{\rm tail, 22}=-4$.

Since we only focus on the $h_{22}$ mode for now, we drop the $(\ell m)$ subscript from the tail terms~\footnote{Resolving tails in higher order modes is more challenging because they decay faster and get overwhelmed by numerical noise quickly. }. For our analysis, we will assume that $h^{\rm QNM}$ consists of only the fundamental mode, and $h^{\rm tail}$ consists of a single power law of the form \eqref{eq:htail} with unknown power-law exponent $p_{\rm tail}$.  The ringdown amplitude of the $h_{22}$ then can be written as: 
\begin{equation}
\label{eq:analytical_model-A22}
\begin{split}
&A_{22}^{\rm ringdown}(t)
=\left[A_{220}^2 e^{-\frac{2t}{\tau_{220}}}+\frac{A_{\rm tail}^2}{(t+c_{\rm tail})^{2p_{\rm tail}}}+\right.\\
&\left.2A_{220}e^{-\frac{t}{\tau_{220}}}A_{\rm tail}(t+c_{\rm tail})^{p_{\rm tail}}\cos\left(\phi_{\rm tail}+\phi_{220}+\omega_{220}t\right)\right]^{1/2}.
\end{split}
\end{equation}

\subsection{Iterative fitting procedure}
\label{sec:fitting_method}
Inspection of Eq. (\ref{eq:analytical_model-A22}) allows us to identify three distinct regimes where each of the three terms in the equation becomes either dominant or non-negligible. For example, in the early times, we expect only the QNM, and therefore the first term ($A_{220}^2 e^{-\frac{2t}{\tau_{220}}}$) in the equation, to be dominant. Similarly, in later times, the QNM amplitude will be extremely small, and the amplitude will mostly consist of the tail terms ($\frac{A_{\rm tail}^2}{(t+c_{\rm tail})^{2p_{\rm tail}}}$). In the intermediate time, both the QNM and tail contributions will be equally important. Mixing between QNM and tail terms will give rise to oscillatory features in the amplitude. This is the regime where the cross terms in the equation cannot be ignored. Furthermore, at even later times, numerical errors in the perturbative framework will start dominating. In Figure~\ref{fig:features}, we show the ringdown amplitude for a binary merger with $[\chi,e]=[0.0, 0.98]$ and highlight all four regimes.

\begin{figure}
\includegraphics[width=\columnwidth]{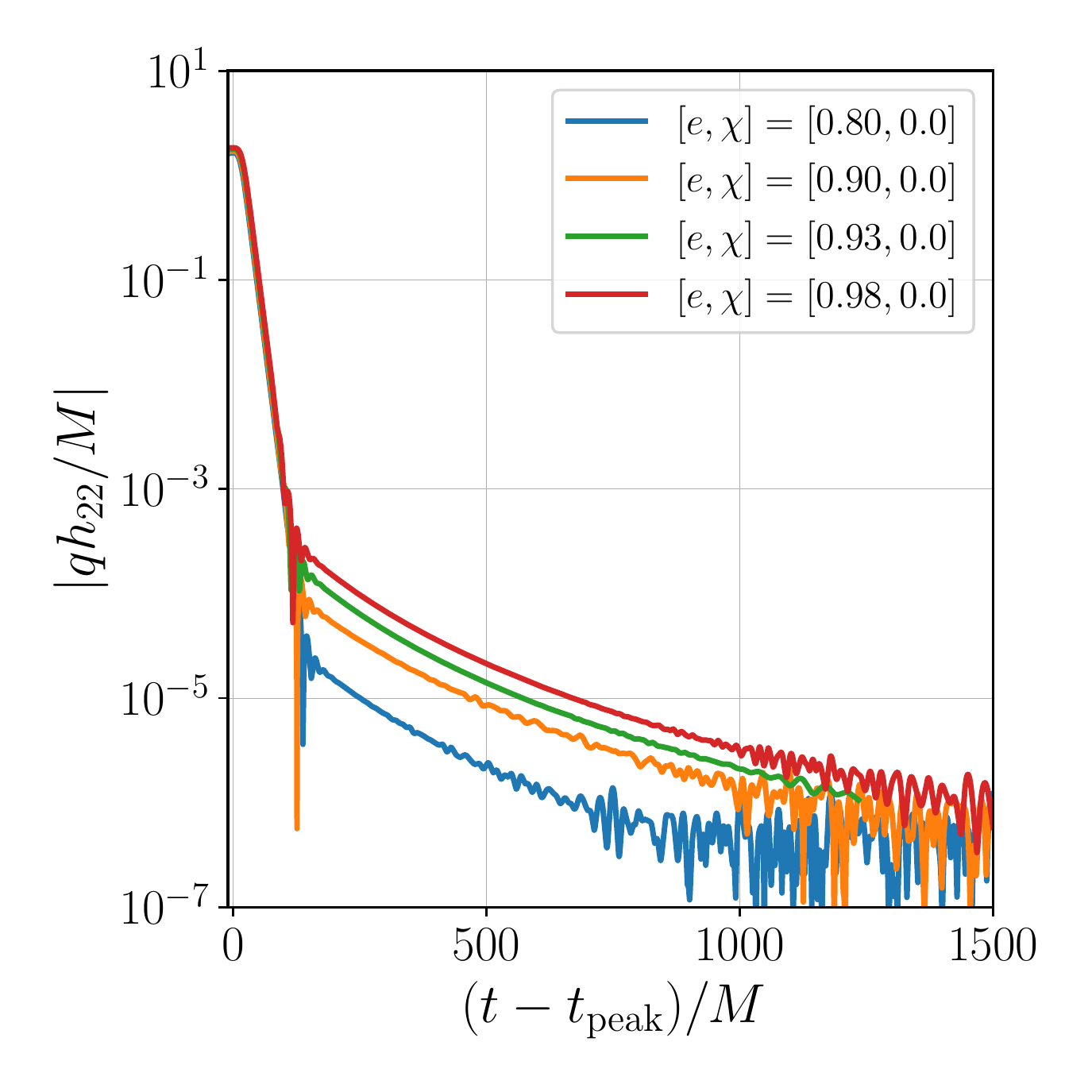}
\caption{We show the tail behavior observed in non-spinning binaries with various eccentricity configurations ranging from $e=0.8$ to $e=0.98$. More details are in Section~\ref{sec:nospin}.}
\label{fig:tails_data}
\end{figure}

We leverage these distinctive features to develop an iterative fitting procedure for the ringdown amplitude. First, we pinpoint the QNM-dominated regime in the ringdown data and exclusively fit it with the QNM amplitude function:
\begin{equation}
A_{22}^{\rm QNM}(t) = A_{220}^2 e^{-\frac{2t}{\tau_{220}}},
\end{equation}
to determine the best-fit values for $A_{220}$ and $\tau_{220}$.
Similarly, we identify the tail-dominated regime and fit the tail amplitude using the function:
\begin{equation}\label{eq:tail}
A_{22}^{\rm tail}(t) = A_{\rm tail}(t+c_{\rm tail})^{p_{\rm tail}}.
\end{equation}
This provides us with the best-fit values for $A_{\rm tail}$, $p_{\rm tail}$ and $c_{\rm tail}$. Now that we have determined four out of the total seven free parameters in Eq. (\ref{eq:analytical_model-A22}), we utilize the intermediate oscillatory data to obtain the remaining two phase parameters ($\phi_{\rm QNM}$, $\phi_{\rm tail}$) and the frequency parameter $\omega_{220}$. This streamlines the fitting procedure, handling fewer free parameters at each step. Note that, while fitting, we truncate ringdown data before it reaches the regime dominated by numerical noise.

\subsection{Software availability}
The extraction of tail parameters involves employing the analytical model described in Section~\ref{sec:model} and the fitting procedure outlined in Section~\ref{sec:fitting_method}. This process is accomplished using the \texttt{gwtails}~\cite{gwtails} Python package, which utilizes \texttt{scipy.curve-fit} in the backend for fitting. Our package is publicly available at \href{https://github.com/tousifislam/gwtails}{https://github.com/tousifislam/gwtails}. The package allows to either perform a combined fit of the QNM and tail contribution or to only fit the tail part.

\begin{figure}
\includegraphics[width=\columnwidth]{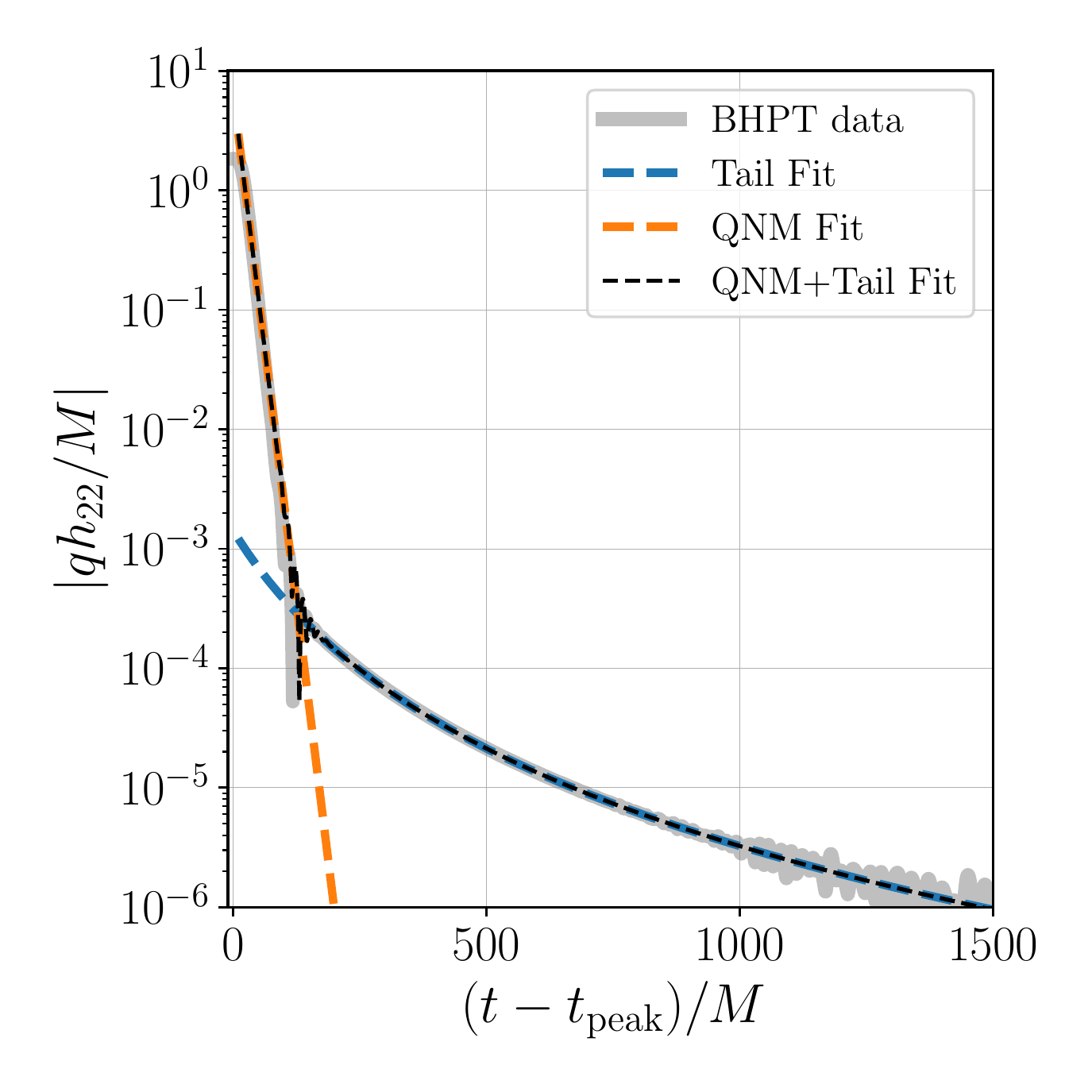}
\caption{We show the $(2,2)$ ringdown amplitude of a non-spinning binary with eccentricity $e=0.98$ (grey solid line). For comparison, we also show the QNM fit in blue dashed line, tail fit in orange dashed line and fit using both QNM and tail in black dashed line. Additionally, we show the fit residuals in the inset. More details are in Section~\ref{sec:nospin}.}
\label{fig:tail_fits_e098}
\end{figure}

\section{Tails in eccentric non-spinning binaries}
\label{sec:nospin}
We simulate eccentric non-spinning BBH mergers with eccentricities ranging from $e=0.8$ to $e=0.98$. 
While the existence of late-time tails has been well understood for many years, recent studies have demonstrated that when the binary has a large eccentricity, these tails become more prominent. Additionally, they emerge earlier than in quasi-circular cases. Note that all our simulations are performed with $\mu=10^{-3}$ but can be easily repeated for other mass ratios. Figure~\ref{fig:tails_data} shows the ringdown amplitude of the $(2,2)$ mode for all eccentric non-spinning binaries. These amplitudes exhibit all four distinct regimes mentioned in Section~\ref{sec:fitting_method}. 
The tail starts occurring mostly around $t=150M$ to $t=200M$. As eccentricity increases, the tail appears earlier, and its amplitude increases. Amplitudes up to $t=100M$ are almost entirely described by the QNM, whereas the tail dominates for $t \ge 200M$. Numerical noise starts dominating from around $t=500M$ for $e=0.8$ and around $t=1000M$ for $e=0.98$.

\begin{figure}
\includegraphics[width=\columnwidth]{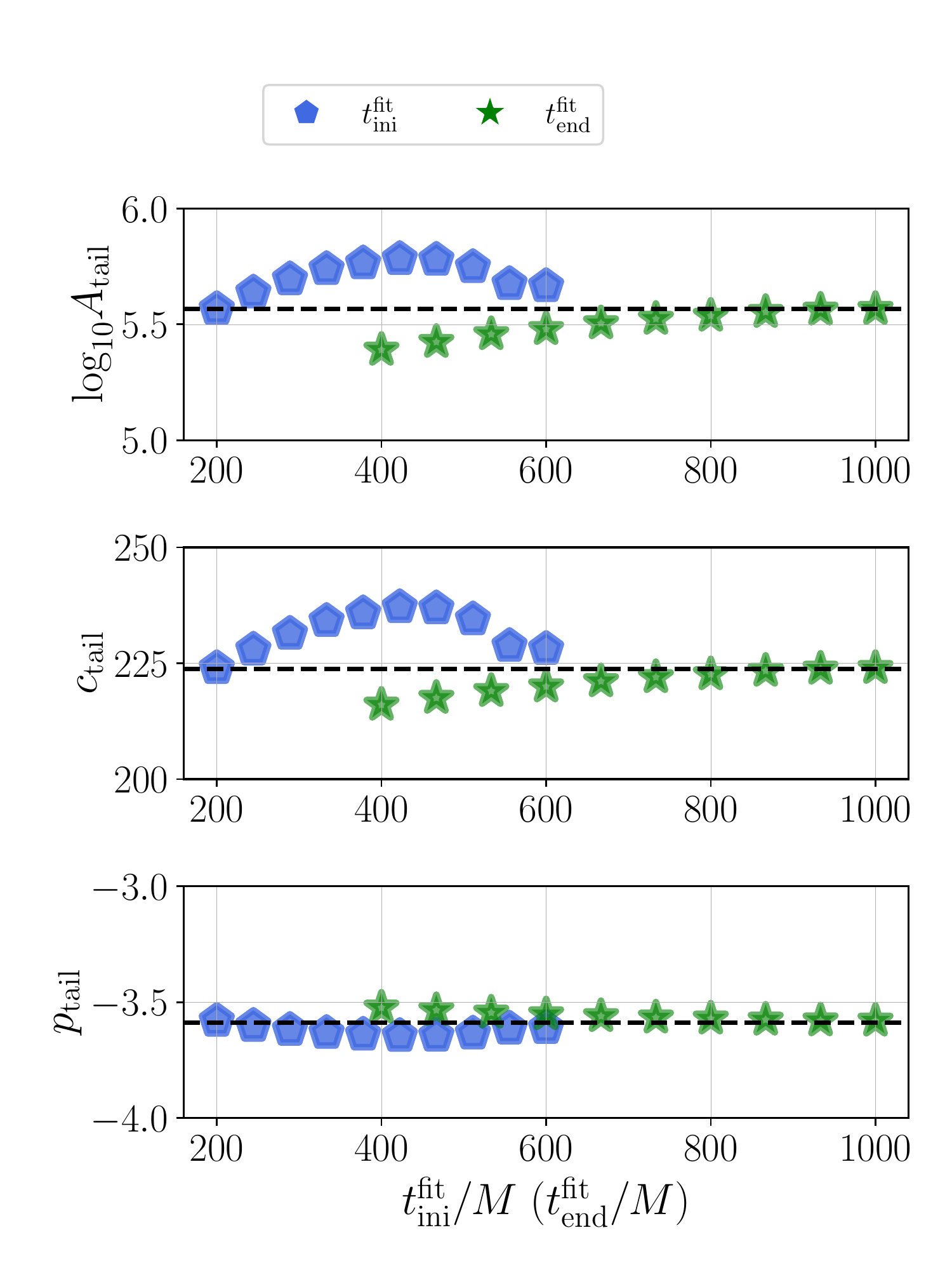}
\caption{We show the extracted best-fit tail parameters ${A_{\rm tail}, c_{\rm tail}, p_{\rm tail}}$ as a function of the initial time (blue pentagons) and as a function of the final time used in the fitting (green stars). In the first case, we fix the final time to be $t=1000M$, while in the latter case, we fix the initial time to be $t=200M$. Black dashed lines denote their values obtained using the full length of the tail data spanning from $t=200M$ to $t=1000M$. More details are in Section~\ref{sec:nospin}.}
\label{fig:robustness}
\end{figure}

\subsection{Fitting the tail}
After identifying different regimes in each of the ringdown data, we apply the iterative fitting approach described in Section~\ref{sec:fitting_method} to extract all post-merger model parameters. We demonstrate our fits for the binary with $e=0.98$. In Figure~\ref{fig:tail_fits_e098}, we show the ringdown data along with the fits. Specifically, we present three fits: (i) only the QNM fit as an orange dashed line (using data within $10M \leq t \leq 70M$), (ii) only the tail fit as a blue dashed line  (using data within $200M \leq t \leq 1000M$), and (iii) a fit using both QNM and tail as a black dashed line~\footnote{Currently, the selection of time intervals for different waveform regimes is performed primarily through visual inspection.}. 
For the tail part, after $t\ge1000M$, numerical noise becomes visually apparent. However, tail fits seem to capture an average trend out of the noisy data. Finally, we combine both QNM and tail terms and provide a complete fit (black dashed line), as explained in Sec.~\ref{sec:fitting_method}, which matches the data in both the QNM and tail regime as well as in the intermediate oscillatory part. 

For the $e=0.98$ system, we find the best-fit values to be $A_{\rm tail} = 3.684 \times 10^{5} \pm 2.376 \times 10^3$, $c_{\rm tail} = 223.96 \pm 0.2$, and $p_{\rm tail} = -3.58 \pm 0.001$ (Fig.~\ref{fig:tail_fits_e098}). Our error bars are computed from the estimated covariance matrix of the best-fit parameters following the procedure described in the \texttt{scipy.optimize.curve\_fit} module documentation. Note that our best-fit value for the tail exponent is closer to the theoretically expected asymptotic value of $-4$ than the values reported in Ref.~\cite{Albanesi:2023bgi}.

\begin{figure}
\includegraphics[width=\columnwidth]{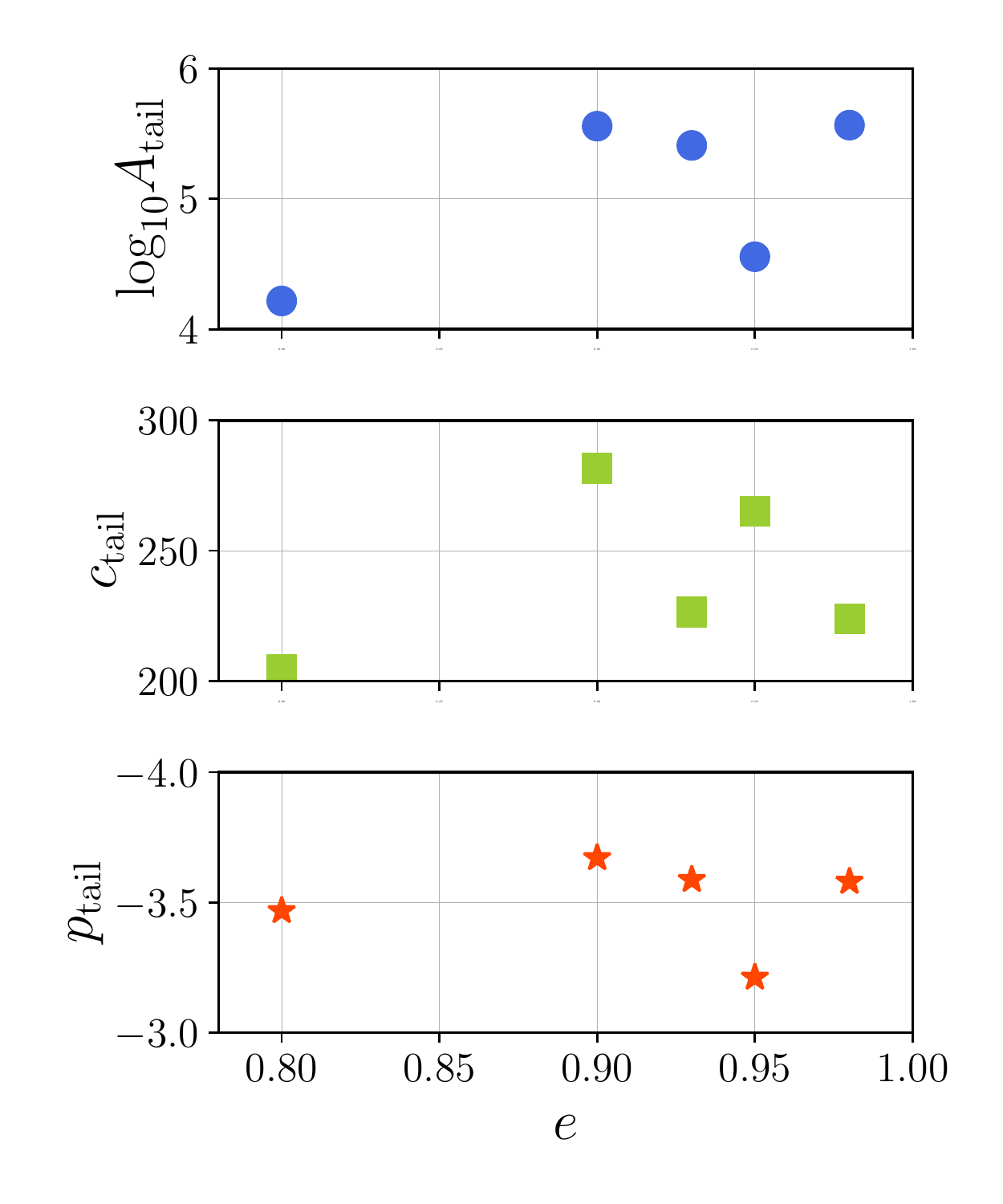}
\caption{We show the extracted best-fit tail parameters ${A_{\rm tail}, c_{\rm tail}, p_{\rm tail}}$ as a function of the eccentricity for binaries with non-spinning configuration. More details are in Section~\ref{sec:nospin}.}
\label{fig:tail_fit_params_nospin}
\end{figure}

Furthermore, for the tail fits, the initial and final times used for the fitting window are $t_{\rm ini}^{\rm fit}=200M$ and $t_{\rm final}^{\rm fit}=1000M$. To verify the robustness of the fits, we repeat our tail fits for varying windows of data. This is done in two ways. First, we fix the final time used in the tail fit to be $t_{\rm final}^{\rm fit}=1000M$ and vary the initial time from $t_{\rm ini}^{\rm fit}=200M$ to $t_{\rm ini}^{\rm fit}=600M$. We have not performed a fit with $t_{\rm ini}^{\rm fit}=800M$ and $t_{\rm final}^{\rm fit}=1000M$ as this choice will lead to using a noise-contaminated short stretch of data covering only $200M$ and result in erroneous best-fit values. Next, we fix the initial time of the fitting window to be $t_{\rm ini}^{\rm fit}=200M$ and vary the final time of the fitting window from $t_{\rm final}^{\rm fit}=400M$ to $t_{\rm final}^{\rm fit}=1000M$. This ensures that the fitting window is \textit{at least} $200M$ long in duration. 
We show the extracted fit parameters in Figure~\ref{fig:robustness} as a function of $t_{\rm ini}^{\rm fit}$ and $t_{\rm final}^{\rm fit}$. We also show the respective best-fit values, obtained using the full length of the tail data spanning from $t=200M$ to $t=1000M$, as a black dashed line. 
We find that changing the fit window does not significantly affect the best-fit values. In particular, changing the initial time of the fitting has a more pronounced effect on the best-fit values than changing the final time of the fitting window. This is because the initial time used in fitting controls the perceived tail amplitude and the time offset $c_{\rm tail}$ in the tail model (see Eq.(\ref{eq:tail})). However, it is noteworthy that the tail exponent $p_{\rm tail}$ is least affected by either the change in the initial or final time in the fitting window. This shows that our estimation of the tail parameters, especially the tail exponent, is robust.

\subsection{Behavior of the tail parameters}
We repeat the fits for all non-spinning eccentric binaries shown in Fig.\ref{fig:tails_data}. As our main focus is understanding late-time tail behavior, we only report tail fits in the rest of the paper.
The results are shown in Fig.\ref{fig:tail_fit_params_nospin}. We observe that the tail amplitude varies between $A_{\rm tail}\sim 10^4$ to $A_{\rm tail}\sim 10^6$, while the time offset parameter varies between $c_{\rm tail}\sim200M$ to $c_{\rm tail}\sim300M$. This is expected, as depending on the eccentricity, the tail features will either get amplified or suppressed, and the time of tail occurrence will change accordingly.
On the other hand, the tail exponent lies between $p_{\rm tail}=-3$ and $p_{\rm tail}=-4$, with most values being close to $p_{\rm tail}=-3.5$. This is close to the expected asymptotic value of $p_{\rm tail}=-4$.

\begin{figure}
\includegraphics[width=\columnwidth]{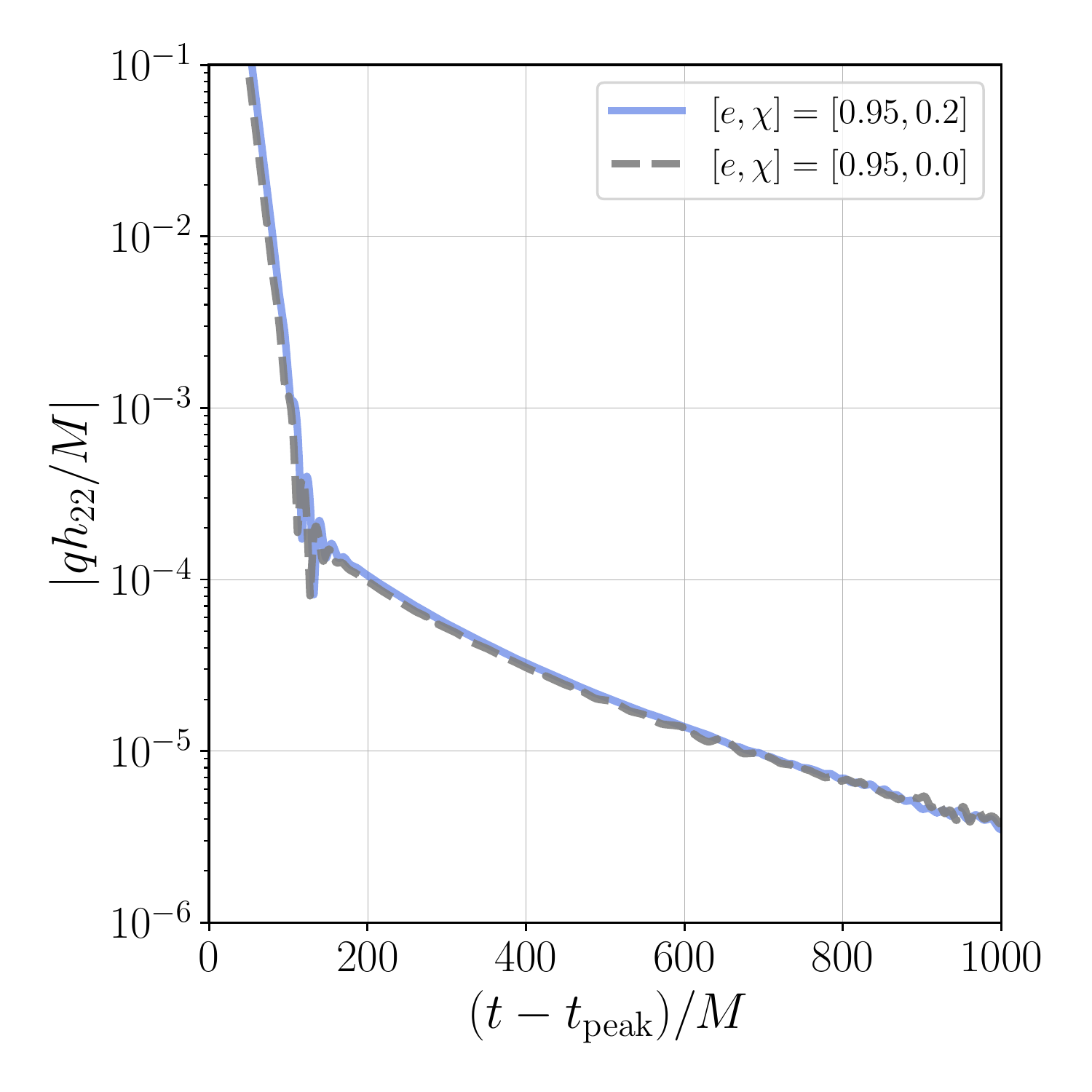}
\caption{We show the $(2,2)$ ringdown amplitude of a spinning binary with $[e,\chi]=[0.95,0.2]$ (blue solid line) and $[e,\chi]=[0.95,0.0]$ (grey dashed line). More details are in Section~\ref{sec:spin}.}
\label{fig:tail_spin_095}
\end{figure}

\begin{figure*}
\includegraphics[width=\textwidth]{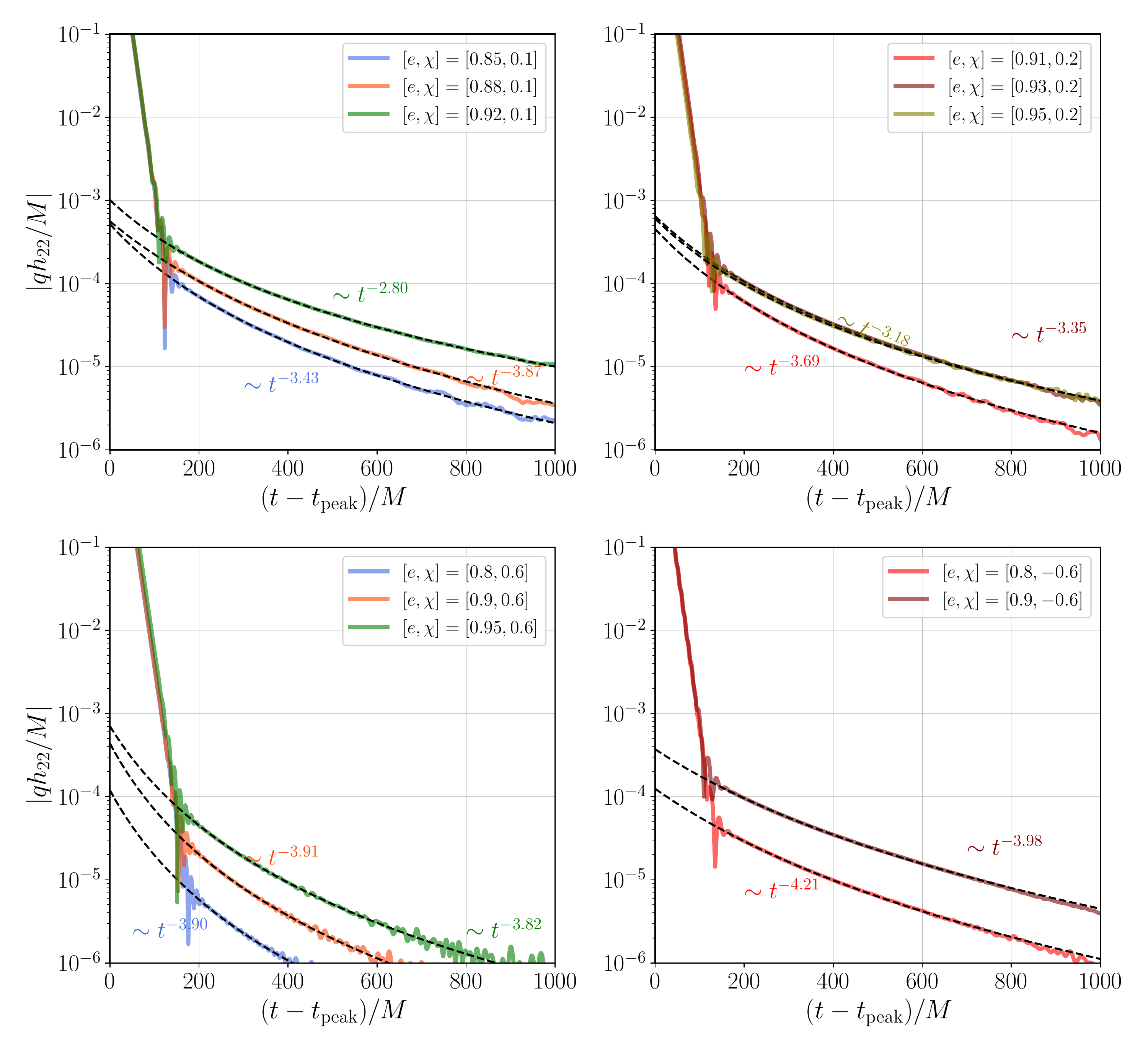}
\caption{We show the $(2,2)$ ringdown amplitude of spinning binaries with varying eccentricity for different spin configurations: $\chi=0.1$ (upper left panel), $\chi=0.2$ (upper right panel), $\chi=0.6$ (lower left panel) and $\chi=-0.6$ (lower right panel). For comparison, we also show corresponding tail fits as black dashed lines. More details are in Section~\ref{sec:spin}.}
\label{fig:tail_spin}
\end{figure*}

\section{Tails in eccentric spinning binaries}
\label{sec:spin}
Next, we proceed to understand whether there is any qualitative change in tail behaviors as we transition from non-spinning to spinning binaries. We simulate a set of mergers where the larger black hole is spinning. 
In particular, we perform four sets of simulations with a dimensionless spin magnitude of $\chi=0.1$, $\chi=0.2$, $\chi=0.6$ and $\chi=-0.6$ for the varying eccentricity configurations. 

In Figure~\ref{fig:tail_spin_095}, we show the ringdown amplitude of the $(2,2)$ mode for a spinning binary with $[e,\chi]=[0.95,0.2]$ (blue solid line) and $[e,\chi]=[0.95,0.0]$ (grey dashed line). We do not find noticeable changes due to the presence of spin. Just like the non-spinning eccentric cases, spinning eccentric binaries also exhibit a fast-decaying QNM regime, an intermediate oscillatory regime, and a late-time tail component.

\begin{figure*}
\includegraphics[width=\textwidth]{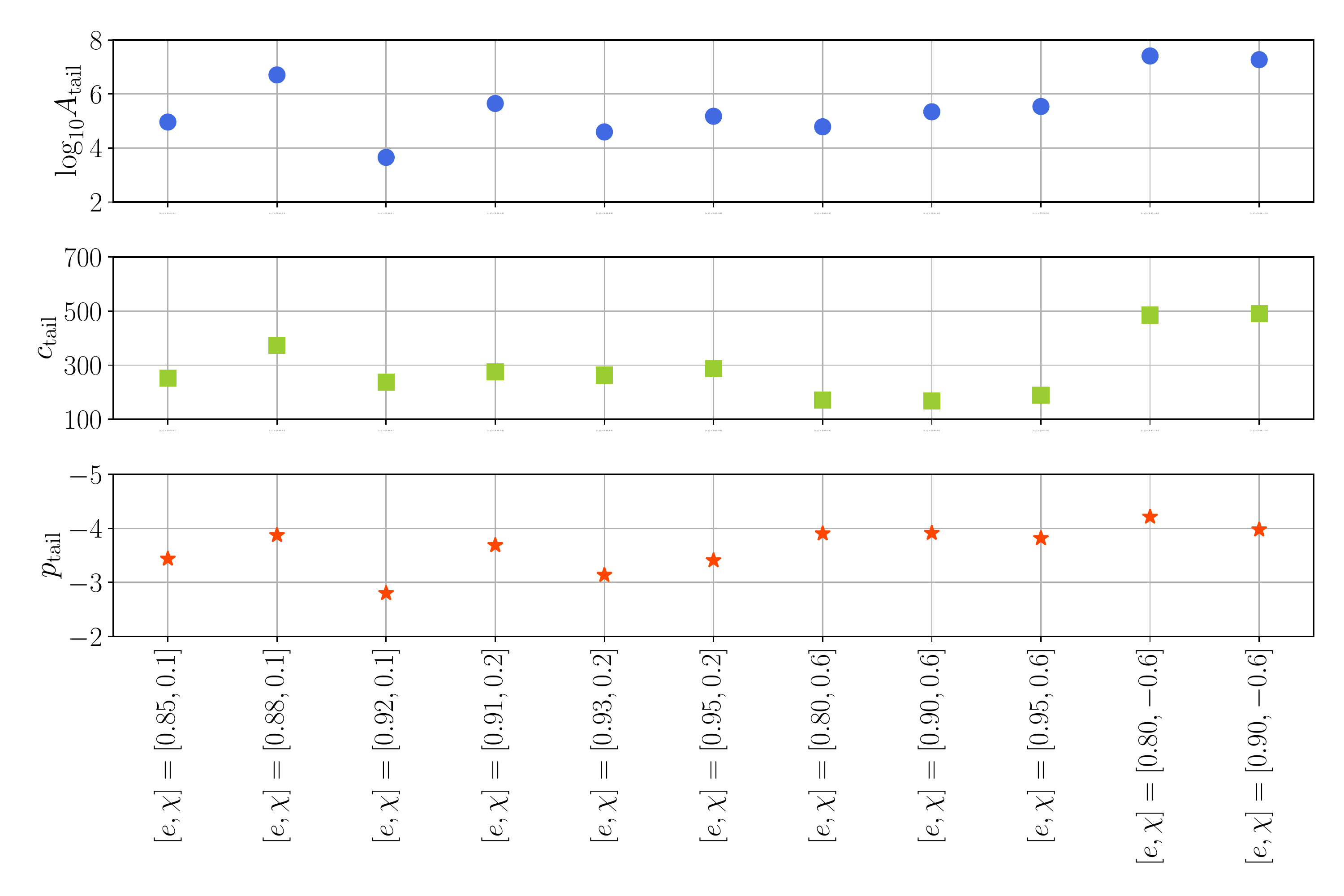}
\caption{We show the extracted best-fit tail parameters ${A_{\rm tail}, c_{\rm tail}, p_{\rm tail}}$ for spinning binaries with different eccentricity configuration. More details are in Section~\ref{sec:nospin}.}
\label{fig:tail_fit_params_spin}
\end{figure*}

We fit these tails with the same power law given in Eq.(\ref{eq:analytical_model-A22}) using \texttt{gwtails} to extract the overall tail behavior. We show the tails and respective fits for all spinning eccentric binaries in Figure~\ref{fig:tail_spin}. We find that the tail model proposed in Eq.(\ref{eq:tail}) still gives a very good fit to the numerical data. Furthermore, just like the non-spinning case, the best-fit value for the tail exponent $p_{\rm tail}$ remains close to the expected asymptotic value of $-4$. Best-fit values for the time-shift parameter $c_{\rm tail}$ lie within $[250M,380M]$ for most cases except for $\chi=-0.6$. This has a significant overlap with the range recovered for the non-spinning case (see Figure~\ref{fig:tail_fit_params_nospin}). For $\chi=-0.6$, time-shift parameter $c_{\rm tail}$ takes a value close to $500M$. 

Below we provide the recovered tail behaviors for different eccentricities and dimensionless spins. 
\begin{subequations}
\begin{align}
[e,\chi] &=[0.85,0.1]: &
A_{22}^{\rm tail} &\sim (t+251.46)^{-3.43}.
\\
[e,\chi] &=[0.88,0.1]: &
A_{22}^{\rm tail} &\sim (t+372.59)^{-3.87}.
\\
[e,\chi] &=[0.92,0.1]: &
A_{22}^{\rm tail} &\sim (t+237.76)^{-2.80}.
\\
[e,\chi] &=[0.91,0.2]: &
A_{22}^{\rm tail} &\sim (t+275.01)^{-3.68}.
\\
[e,\chi] &=[0.93,0.2]: &
A_{22}^{\rm tail} &\sim (t+263.01)^{-3.13}.
\\
[e,\chi] &=[0.95,0.2] &
A_{22}^{\rm tail} &\sim (t+287.21)^{-3.41}.
\\
[e,\chi] &=[0.8,0.6]: &
A_{22}^{\rm tail} &\sim (t+170.35)^{-3.90}.
\\
[e,\chi] &=[0.9,0.6]: &
A_{22}^{\rm tail} &\sim (t+167.62)^{-3.91}.
\\
[e,\chi]&=[0.95,0.6]: &
A_{22}^{\rm tail} &\sim (t+189.25)^{-3.81}.
\\
[e,\chi]&=[0.8,-0.6]:&
A_{22}^{\rm tail} &\sim (t+484.99)^{-4.21}.
\\
[e,\chi]&=[0.9,-0.6]:&
A_{22}^{\rm tail} &\sim (t+491.19)^{-3.98}.
\end{align}
\end{subequations}
Extracted best-fit tail parameters ${A_{\rm tail}, c_{\rm tail}, p_{\rm tail}}$ for spinning binaries with different eccentricity configurations are shown in Figure~\ref{fig:tail_fit_params_spin}.

\section{Understanding the source of tails}
\label{sec:tail_reason}
To better understand the late-time tail behavior observed in eccentric BBH mergers, we perform a series of numerical experiments to identify the specific characteristics in a BBH evolution that excite late-time tails more strongly.

\begin{figure}
\includegraphics[width=\columnwidth]{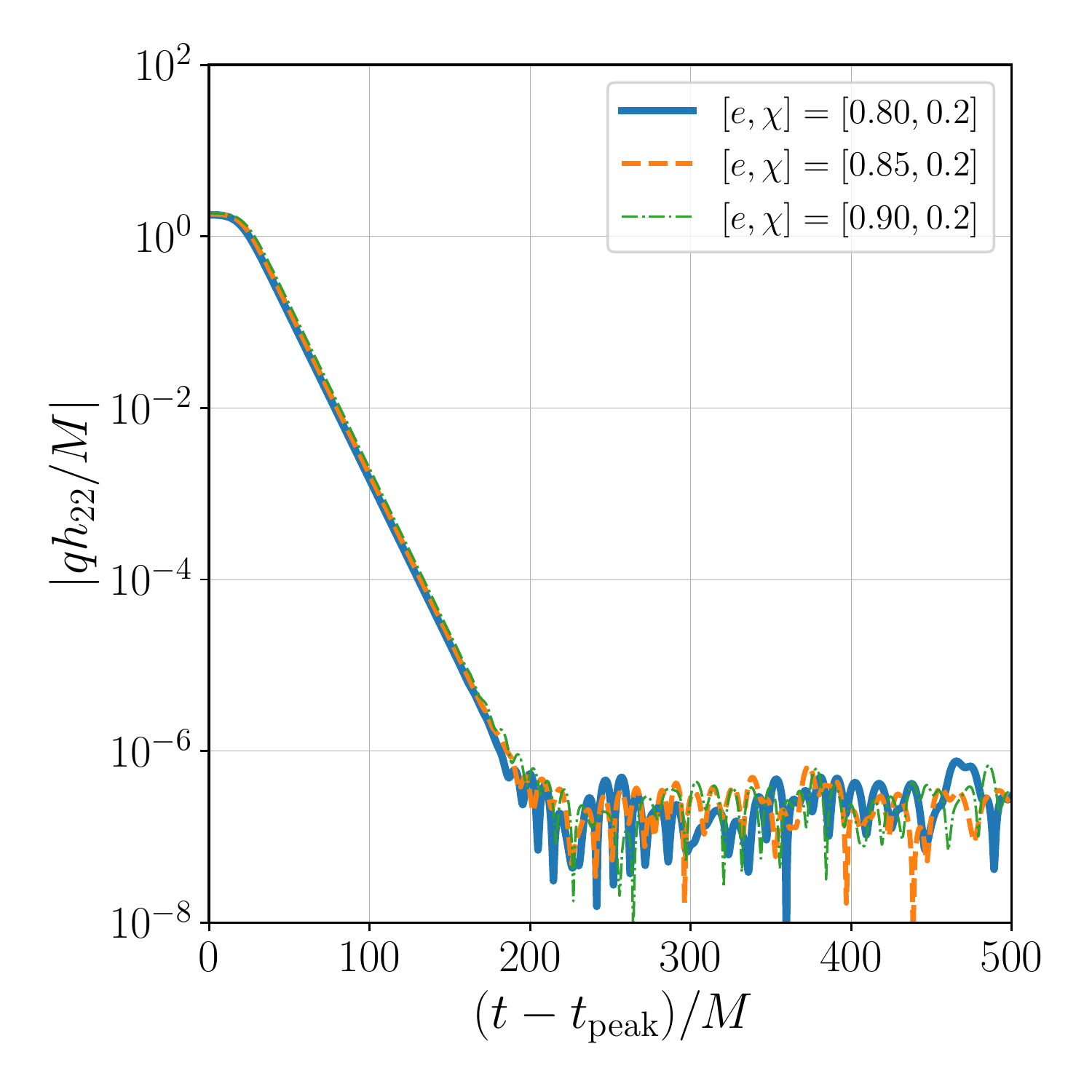}
\caption{We show the $(2,2)$ ringdown amplitude for binaries on a geodesic plunge that asymptotically starts at the last stable orbit with spin $\chi=0.6$ and eccentricities $e=[0.8, 0.85, 0.9]$. We find no evidence of tails within the resolution of our numerical simulation. More details are in Section~\ref{sec:tail_reason}.}
\label{fig:tail_geodesic}
\end{figure}

We first replace EOB trajectories (that incorporate radiation-reaction) with geodesic plunge orbits utilizing the closed-form solutions given in Ref.~\cite{Dyson:2023fws}. We use the \texttt{KerrGeodesics}~\cite{KerrGeodesics} and \texttt{KerrGeoPy}~\cite{KerrGeoPy} packages in the Black Hole Perturbation Toolkit~\cite{BHPToolkit} to obtain these solutions. We start the simulation at the LSO. We then compute the waveform following the same procedure described in Section~\ref{sec:wave_gen}. In Figure~\ref{fig:tail_geodesic}, we show the ringdown amplitude of the $(2,2)$ mode for binaries following a plunge geodesic that is very close to the LSO with spin $\chi=0.6$ and eccentricities at the LSO $e=[0.8, 0.85, 0.9]$~\footnote{We present the respective trajectories in Section~\ref{sec:geo_trajectory} and in Figure~\ref{fig:trajs_geo_plot}.}. We find that eccentricity does not noticeably alter the amplitudes in these cases. Furthermore, amplitudes decrease monotonically (QNM decay) until they reach the numerical noise floor at $\sim 200M$. We find no sign of tails. Note that there are almost seven orders of magnitude difference between the peak amplitude at $t=0$ and the noise floor. On the other hand, peak amplitudes and tail amplitudes at the beginning differ by mostly four to five orders of magnitudes (see Figure~\ref{fig:tails_data}). This suggests that tails are not generated in the final stage of the plunge.

\begin{figure}
\includegraphics[width=\columnwidth]{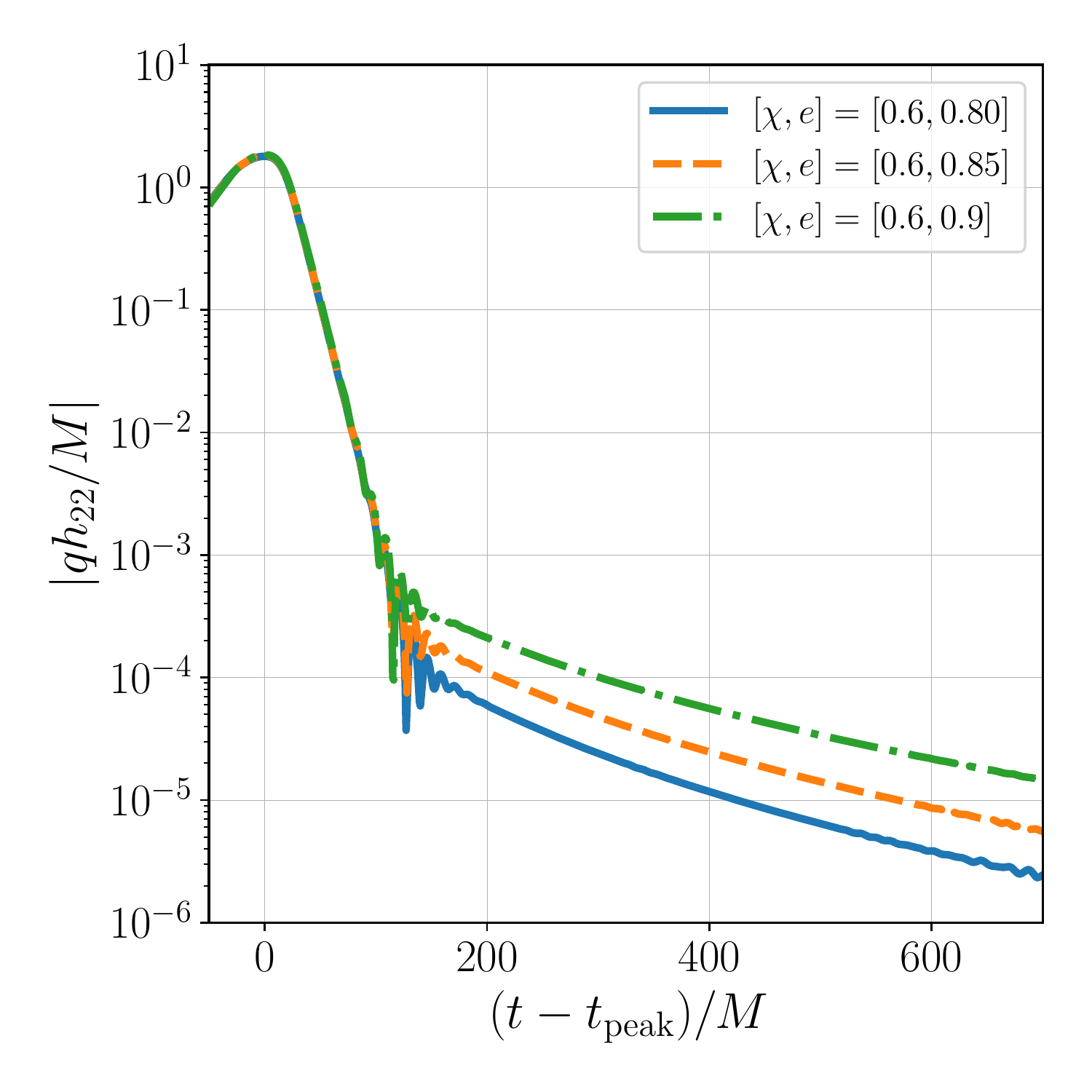}
\caption{We show the $(2,2)$ ringdown amplitude for binaries on a geodesic plunge with spin $\chi=0.6$ and eccentricities $e=[0.8, 0.85, 0.9]$. We start the simulation at the last apocenter passage before the last stable orbit. Compared to Fig.~\ref{fig:tail_geodesic}, where no tails were found, this longer orbit excites late-time tails. More details are in Section~\ref{sec:tail_reason}.}
\label{fig:tail_long_geodesic}
\end{figure}

Next, we consider geodesic plunging trajectories that start from the last apocenter passage, i.e. plunge geodesics with their sole radial turning point outside the LSO radius for the corresponding angular momentum. These trajectories have an energy slightly larger than the LSO energy and they do not manifest whirls effect around the LSO radius. We find that these orbits produce late-time tails (e.g. Fig.~\ref{fig:tail_long_geodesic}; for eccentricity at the LSO $e=0.9$ and spin $\chi=0.6$).

\begin{figure}
\includegraphics[width=\columnwidth]{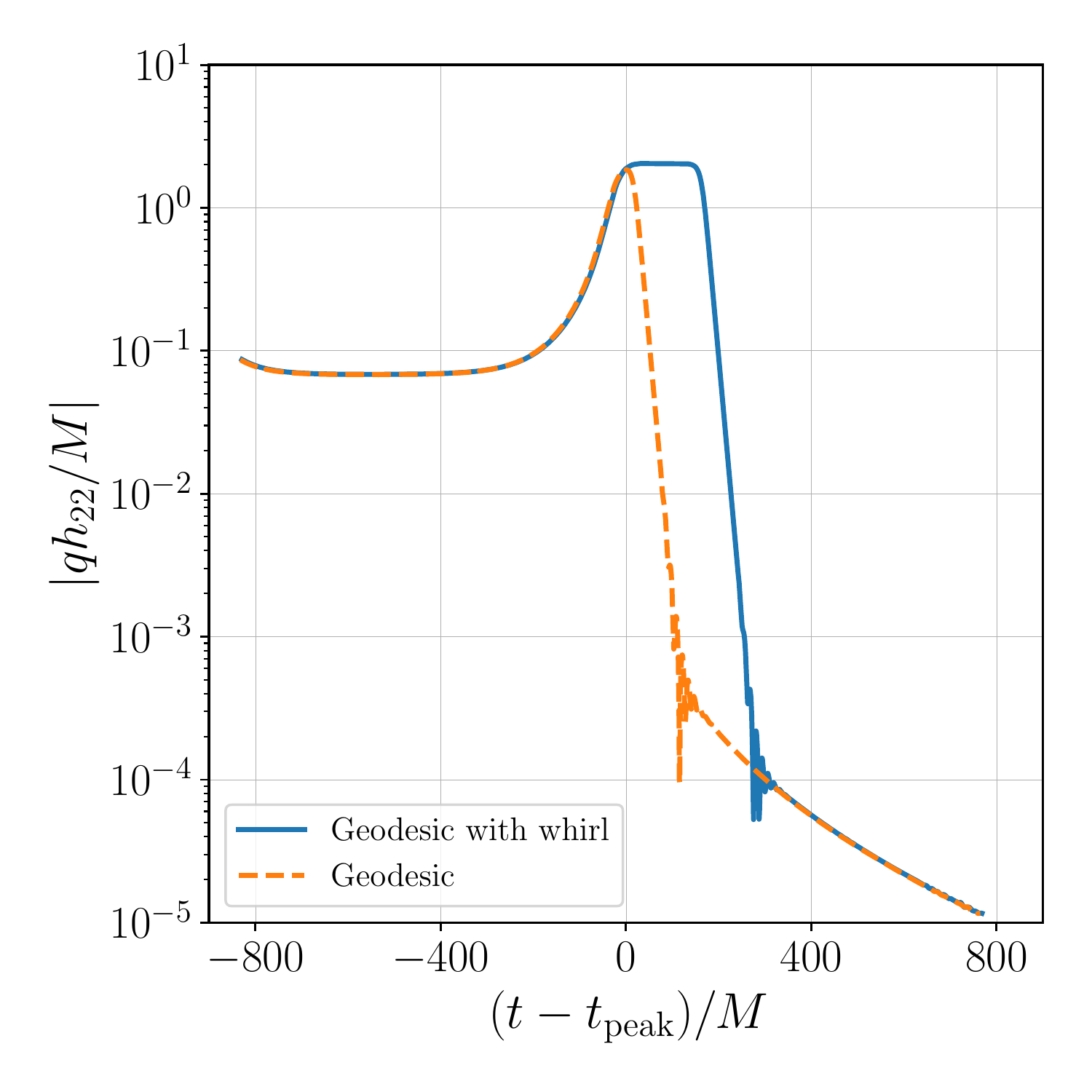}
\caption{We show the $(2,2)$ ringdown amplitude for binaries on a geodesic plunge with spin $\chi=0.6$ and eccentricity $e=0.9$. We start both simulations at the last apocenter. One of the simulations has delayed plunge. However, both orbits generate consistent tail behavior. More details are in Section~\ref{sec:tail_reason}.}
\label{fig:whirl}
\end{figure}

The two sets of plunging geodesics mentioned so far start at different locations. The trajectories of Fig.~\ref{fig:tail_geodesic} start at the LSO radius in the asymptotic past, while the trajectories of Fig.~\ref{fig:tail_long_geodesic} start at the last apocenter passage and do not whirl long on the LSO radius. This may suggest that either the absence of an apocenter passage or the presence of circular whirls at the LSO radius may affect the tail excitation.
In order to assess this last point, we simulate two eccentric BBH mergers from the last apocenter with spin $\chi=0.6$ and LSO eccentricity $e=0.9$. We fine tune the energy and the angular momentum of these orbits so that they start from the same last apocenter radius and have similar evolution up to the LSO radius. At this radial point, one of the orbits includes whirls around the LSO radius before the merger, while the other orbit does not. In Fig~\ref{fig:whirl} we show the simulated amplitudes of the $(2,2)$ mode of these two orbits, aligned at the starting time (the same last apocenter passage). As expected, the presence of whirls will generate a delayed merger. Interestingly, in both cases, we observe tails and the tails matchup almost perfectly not only in terms of exponents, but also in terms of amplitude and delay. This suggests that the generation of the peak of the waveforms (and the QNMs that follow) is decoupled from the generation of the tails. It further suggests that the generation of the tails is linked to the part of the trajectory where the two trajectories are close, i.e. the period just after the last apocenter passage.
 
\begin{figure}
\includegraphics[width=\columnwidth]{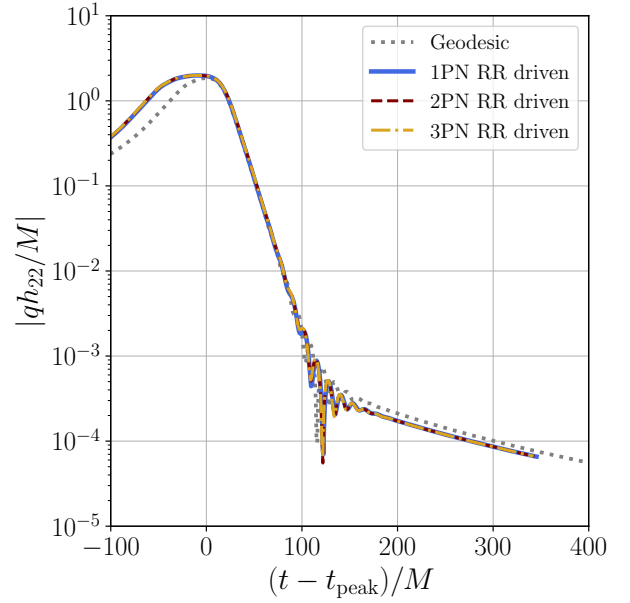}
\caption{We show the $(2,2)$ ringdown amplitude for binaries (with spin $\chi=0.6$ and eccentricity $e=0.9$) on evolving orbits driven by radiation reaction (RR) forces calculated upto 1PN (blue solid line), 2PN (marron dashed line) and 3PN (yellow dotted line) orders. For comparison, we also show the tails observed in a geodesic plunge orbit (grey line). Waveforms are aligned at the apocenter for visual clarity. More details are in Section~\ref{sec:tail_reason}.}
\label{fig:PN}
\end{figure}

Overall, Figs.~\ref{fig:tail_geodesic} and \ref{fig:tail_long_geodesic} suggest that the emergence of late-time tails in the waveform does not strictly require orbital evolution driven by radiation reaction. However, certain quantitative aspects—such as the tail amplitudes and power-law decay exponents—might exhibit slight differences between waveforms generated by radiation-reaction-driven orbits and those produced from geodesic plunges. We next examine this possibility. In Fig.~\ref{fig:PN}, we present the $(2,2)$ mode amplitudes of waveforms generated by three trajectories evolved using RR forces truncated at 1PN, 2PN, and 3PN orders, all with similar apocenter passages. These trajectories have a fixed spin parameter of $\chi=0.6$ and an LSO eccentricity of $e=0.9$. For comparison, we also include the amplitude of a waveform from a plunging orbit with a similar apocenter. We observe that the different orders of RR force yield similar tails. Moreover, all waveforms appear to be qualitatively consistent with the geodesic tail.

At this point, we emphasize that the results presented here correspond to a spin value of $\chi = 0.6$. However, we have verified that the qualitative features remain consistent across other values of spin as well. The results of this section suggest that {\em late-time tails are strongly excited in scenarios wherein the particle-source of the Teukolsky equation is localized far from the black hole, i.e. in the neighborhood of the apocenter on a highly eccentric orbit.} 

A similar conclusion has been recently reached in an independent investigation using RWZ formalism~\cite{DeAmicis:2024not}. Given that {\em the tails are a low-frequency (long-wavelength) phenomenon -- they arise from the branch-cut in the Greens function on the imaginary axis -- it is reasonable to expect that their amplitude would be impacted by low-frequency perturbations of the type that would be sourced by large radius orbital motion}. See the Appendix~\ref{app:model} for further details on this point. 

\section{Hints of tails from numerical relativity}
\label{sec:NR}
Previously, Ref.~\cite{Carullo:2023tff} have reported indications of late-time tails in non-spinning eccentric RIT-NR simulations. It is important to note that these simulations have a relatively shorter duration, reaching only up to approximately $\sim 150M$ after the merger. Given the current limitations of NR simulations, which do not extend far into the ringdown regime, identifying precise tail behavior in the data remains challenging. 

\begin{figure}
\includegraphics[width=\columnwidth]{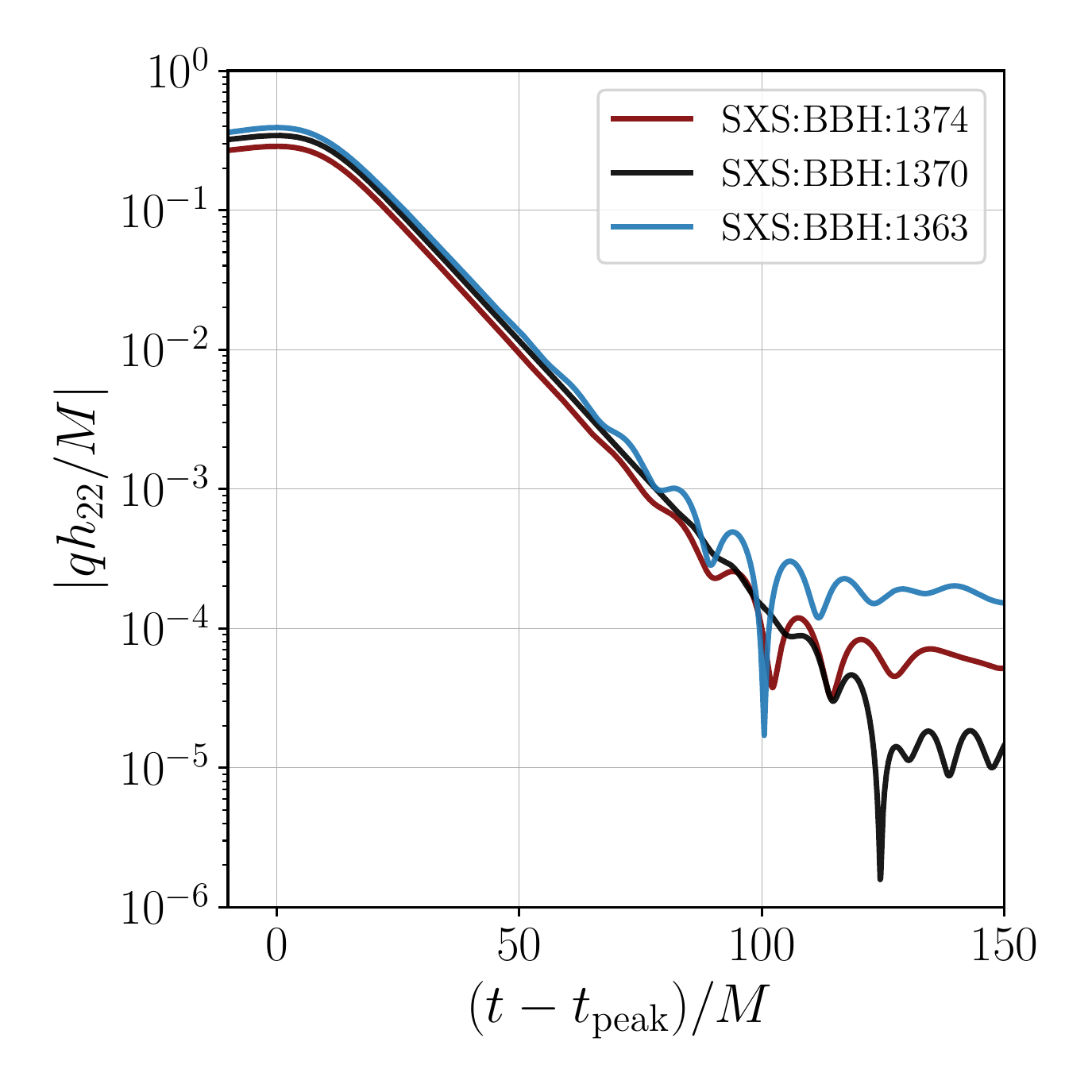}
\caption{We observe an intermediate oscillatory regime in three non-spinning SXS-NR simulations: \texttt{SXS:BBH:1363}, \texttt{SXS:BBH:1370}, and \texttt{SXS:BBH:1374}. These simulations are characterized by the following parameter values: $[q,e]=[1, 0.19]$, $[q,e]=[2, 0.19]$, and $[q,e]=[3, 0.18]$. We have checked that these features appear at different levels of numerical resolution and waveform extrapolation order. These oscillations indicate the waveform is transitioning away from exponential decay, which, in the perturbation theory calculation, always precedes the onset of tails. However, without conclusive evidence of tails, we cannot rule out many possible alternative explanations.
More details are in Section~\ref{sec:NR}.}
\label{fig:SXSNR}
\end{figure}

We have examined publicly available NR simulations from the SXS collaboration and found features similar to those reported in Ref.~\cite{Carullo:2023tff}, which are potentially suggestive of Price tails. In Figure~\ref{fig:SXSNR}, we show the $(2,2)$ mode amplitude for three representative non-spinning SXS-NR simulations with high eccentricities ($e \sim 0.19$ as estimated at a dimensionless reference frequency of $x=0.075$). These simulations correspond to mass ratios of $q=1$, $q=2$, and $q=3$, respectively. While these NR waveforms do not explicitly and convincingly show the tail part, they exhibit an oscillatory intermediate regime that always proceeds the onset of tails; cf. Fig.~\ref{fig:features}. Similar features were reported using the RIT NR data~\cite{Carullo:2023tff}. 
Yet, unlike the RIT waveforms, the SXS waveforms appear to have the transition from exponential decay to intermediate oscillatory behavior about where we would expect it to be based on perturbation theory: about $4$ orders of magnitude smaller than the peak at about $t\approx100M$. But this still does not conclusively identify what it is. While we have checked these features are similar at different levels of numerical resolution and waveform extrapolation order, other small systematic effects (e.g. boundary conditions, a small piece of GW memory, or something else) could be responsible for the observed behavior. 

At this point, it is important to note that there are only a handful of eccentric NR simulations publicly available. 
Moreover, current NR simulations do not extend into the proper tail regime yet. This limitation currently prevents us from making a direct apples-to-apples comparison between NR and BHPT tail behaviors. However, as more data becomes available, we anticipate performing such a systematic and comprehensive comparison in the future.
Indeed, recent studies completed during the journal review
of our paper have, for the first time, unambiguously computed late-time
gravitational-wave tails using fully nonlinear, 3+1 dimensional
numerical relativity simulations of merging black holes~\cite{DeAmicis:2024eoy,Ma:2024hzq}.

\begin{figure}
\includegraphics[width=\columnwidth]{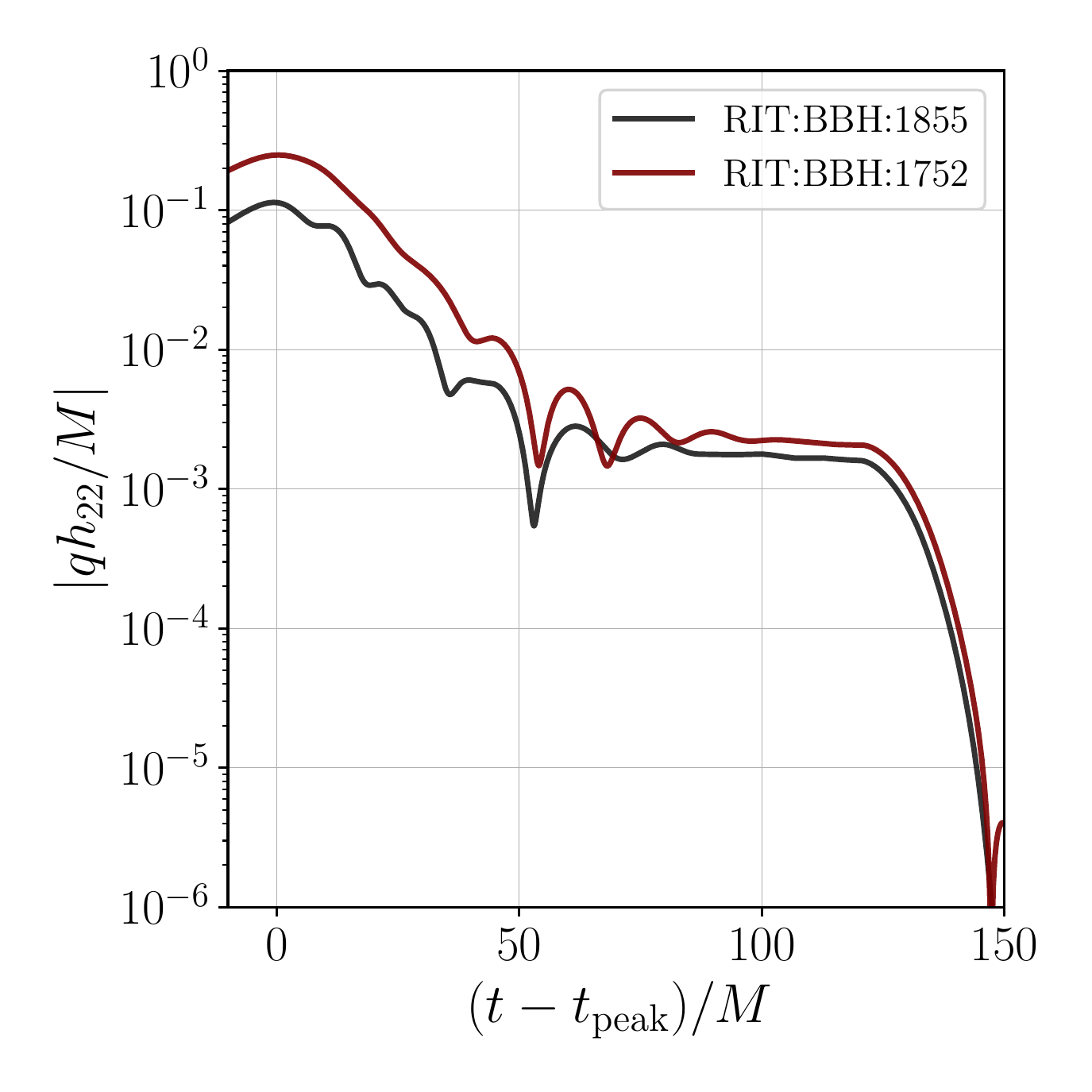}
\caption{We observe an intermediate oscillatory regime in two spinning RIT-NR simulations: \texttt{RIT:BBH:1363} and \texttt{RIT:BBH:1855}. These simulations are characterized by the following parameter values: $[q,e,\chi]=[4, 0.91, -0.81]$, $[q,e,\chi]=[1, 0.79, -0.8]$. More details are in Section~\ref{sec:NR}. These oscillations indicate the waveform is transitioning away from exponential decay, which, in the perturbation theory calculation, always precedes the onset of tails. However, without conclusive evidence of tails, we cannot rule out many possible alternative explanations.
More details are in Section~\ref{sec:NR}.}
\label{fig:RITNR}
\end{figure}

Next, we examine a set of RIT-NR simulations for eccentric spinning binaries.
In Figure~\ref{fig:RITNR}, we show the ringdown amplitude for two spinning eccentric binaries. These simulations are characterized by the following parameter values: $[q,e,\chi]=[4, 0.91, -0.81]$, and $[q,e,\chi]=[1, 0.79, -0.8]$. While the first binary has only the larger black hole spinning, for the latter, both black holes are spinning. We observe a sudden drop in amplitude right after the oscillatory part, likely due to the numerical resolution limit in these NR simulations.

\section{Discussion and conclusion}
In this paper, we employ black hole perturbation theory, within the Teukolsky equation framework, to investigate the Price tails in eccentric binary black hole mergers. Our study reveals that the presence of eccentricity amplifies the effects of tails in the late-time evolution of BBH mergers. This corroborates findings from previous works~\cite{Albanesi:2023bgi, Carullo:2023tff, DeAmicis:2024not}, which utilized perturbative RWZ framework in BBH simulations and NR, respectively.

We demonstrate that the eccentricity-induced slowly-decaying tails in non-spinning BBH mergers, as predicted by BHPT data, closely adhere to their expected asymptotic behavior. A notable advancement in our study involves the examination of spinning eccentric binaries, which follow tail behavior similar to that observed in non-spinning eccentric cases. Furthermore, we introduce an efficient framework for identifying various qualitative regimes in the late-time tail evolution and fitting the tail behavior with an analytical model. The robustness of our fitting method is explored and found to be reliable. Finally, we investigate the dependence of the best-fit model parameters on the spin and eccentricity values of the binary. 

While our results support the existence (and enhancement) of tails in eccentric BBH mergers as reported in Ref.~\cite{Albanesi:2023bgi,Carullo:2023tff}, we find that the decay rate of the tails in both non-spinning and spinning eccentric binaries lies between $-4$ and $-3$ instead of $-1.3$ as found in Ref.~\cite{Albanesi:2023bgi} or in between $-3$ and $-2$ as observed in Ref.~\cite{Carullo:2023tff}. Our recovered values are therefore closer to the expected value $-(\ell+2)$ (i.e. $-4$ for the $\ell=2$ multipole considered here) than Ref.~\cite{Albanesi:2023bgi} and Ref.~\cite{Carullo:2023tff}. We note that, due to the shorter length of post-merger NR data, Ref.~\cite{Carullo:2023tff} could only analyze gravitational waves up to $\sim 100M$ after the merger. Ref.~\cite{Albanesi:2023bgi} however has evolved the system up to $\sim 300M$ after merger. On the other hand, our simulations extend up to $\sim 1000M$ after the merger or beyond. This gives us a unique opportunity to probe the late-time tails more robustly.

We also offer compelling evidence for the fact that the late-time tails (or Price tails) are strongly excited in eccentric BBH systems when the secondary is in the neighborhood of the apocenter of the eccentric orbit, as opposed to any structure in the strong field (eg. LSO, peak of the potential, photon sphere, etc.) of the primary. This is because perturbations sourced in that manner are low-frequency and that is key to the excitation of strong amplitude tails. Appendix.~\ref{app:model} provides further intuition and evidence on this point for both orbital motion and wave propagation. These studies complement the analysis performed in Ref.~\cite{Cardoso:2024jme} to understand the origin of pronounced tails using second-order perturbation and dynamical settings.

While our work offers a more intricate exploration of the phenomenology of tails in eccentric BBH mergers, certain questions remain. For instance, it would be valuable to empirically confirm the decay rate computed for these cases will eventually reach its expected asymptotic value of $p_{\rm tail}=-4$. Addressing this would necessitate extending the simulation well beyond our current final time, but our current code resolution of $10^{-6}$-$10^{-5}$ (cf. Figure 6 of Ref.~\cite{Islam:2022laz} and Figure 5 of Ref.~\cite{Rink:2024swg}) is insufficient for such scenarios. Future efforts, with the availability of higher-order black hole perturbation theory (BHPT) codes, may provide insights into these unresolved questions.

Certainly, exploring the systematic behavior of tail contributions across a wide range of binary parameters, including mass ratio, eccentricity, and spins, holds significant value. Such an investigation could contribute to the development of an efficient analytical model for tail contributions as well as their impact on data analysis efforts. We leave this for future work.

Just before the completion of this manuscript, the paper by De Amicis \textit{et al.}~\cite{DeAmicis:2024not} appeared on the arXiv. The two analyses were conducted independently and offer complementary perspectives on the phenomenology and origin of late-time tails in merging eccentric binaries. While De Amicis et al. focused solely on radiation-reaction driven orbits, our study examined both radiation-reaction driven and geodesic orbits. Furthermore, we explored both a Schwarzschild and Kerr cases while De Amicis \textit{et al.} investigated only Schwarzschild cases. De Amicis et al.\cite{DeAmicis:2024not} provided an analytical model for the observed tail behavior in terms of the system source and Green's function, whereas our study employed numerical approaches to understand the origin of these late-time tails. Both studies concluded that these tails are strongly excited in eccentric BBH systems, particularly when the smaller black hole is near apocenter.

\begin{acknowledgments}
We thank Vijay Varma and Gregorio Carullo for helpful discussions and thoughtful comments on the manuscript. We also thank the SXS collaboration and RIT gravity group for maintaining a publicly available catalog of NR simulations that has been used in this study. 
Part of this work is additionally supported by the Heising-Simons Foundation, the Simons Foundation, and NSF Grants Nos. PHY-1748958. S.E.F and G.K. acknowledge support from NSF Grant No. DMS-2309609. G.K. acknowledges support from NSF Grant No. PHY-2307236. S.E.F acknowledges support from NSF Grant No. PHY-2110496. Simulations were performed on CARNiE at the Center for Scientific Computing and Visualization Research (CSCVR) of UMassD, which is supported by the ONR/DURIP Grant No.\ N00014181255 and the UMass-URI UNITY HPC/AI supercomputer supported by the Massachusetts Green High Performance Computing Center (MGHPCC). 
This work makes use of the Black Hole Perturbation Toolkit~\cite{BHPToolkit}.
\end{acknowledgments}

\appendix
\section{Examples of tail generation and excitation}
\label{app:model}

In this Appendix, we provide some intuition behind tail excitation by considering two examples. We empirically show that tails are more strongly excited for orbits and waves with lower frequency content. Further insight is obtained by considering the structure of near-field-to-far-field waveform propagation kernels. Throughout this appendix, we set the larger black hole's mass to $M=1$.

\begin{figure}
\includegraphics[width=\columnwidth]{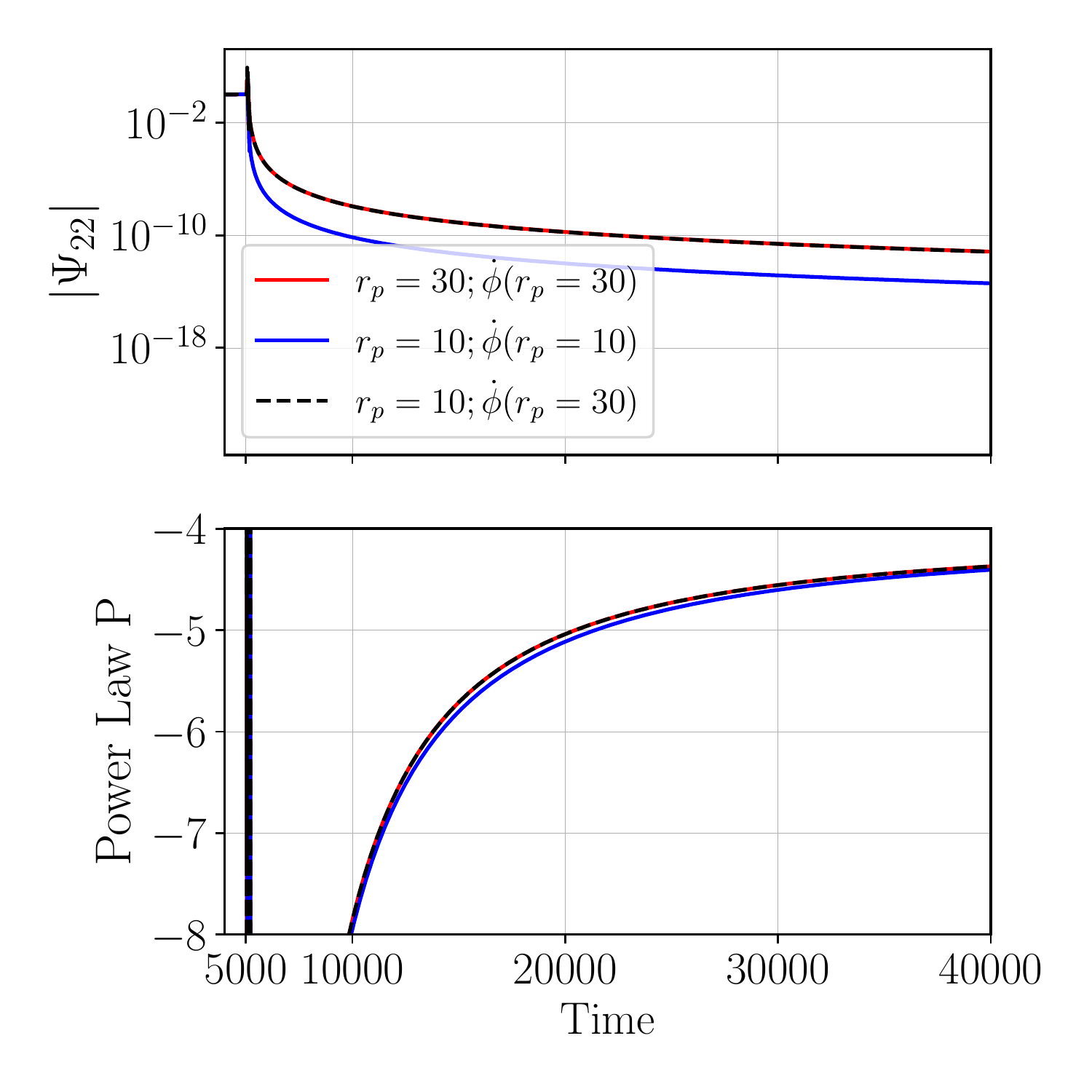}
\caption{The (normalized) Zerilli function  $\left| \Psi_{22} \right| $ as observed at future null infinity for three different circular orbits: a geodesic orbit where the smaller black hole is located at $r_p =10$ (solid blue line), a geodesic orbit where the smaller black hole is located at $r_p =30$ (solid red line), and a non-geodesic orbit where the smaller black hole is located at $r_p =10$ but the orbital frequency is set to match that of the $r_p=30$ geodesic orbit (dashed black line). In all cases, the source terms are smoothly turned off to observe the tail excitation driven by this circular orbit. We find that tail excitation is predominately driven by the orbital frequency of the problem as opposed to the secondary black hole's radial location, and problems with lower frequencies lead to larger tails. The bottom figure shows tail decay rates, $P$, for $\Psi \propto t^{-p}$ computed using logarithmic difference quotients. The decay rate $P=-4$ is the theoretical prediction for $\ell=2$ waves at future null infinity~\cite{Gundlach:1993tp}, which we are slowly approaching.}
\label{fig:AppA}
\end{figure}

\begin{figure}
\includegraphics[width=\columnwidth]{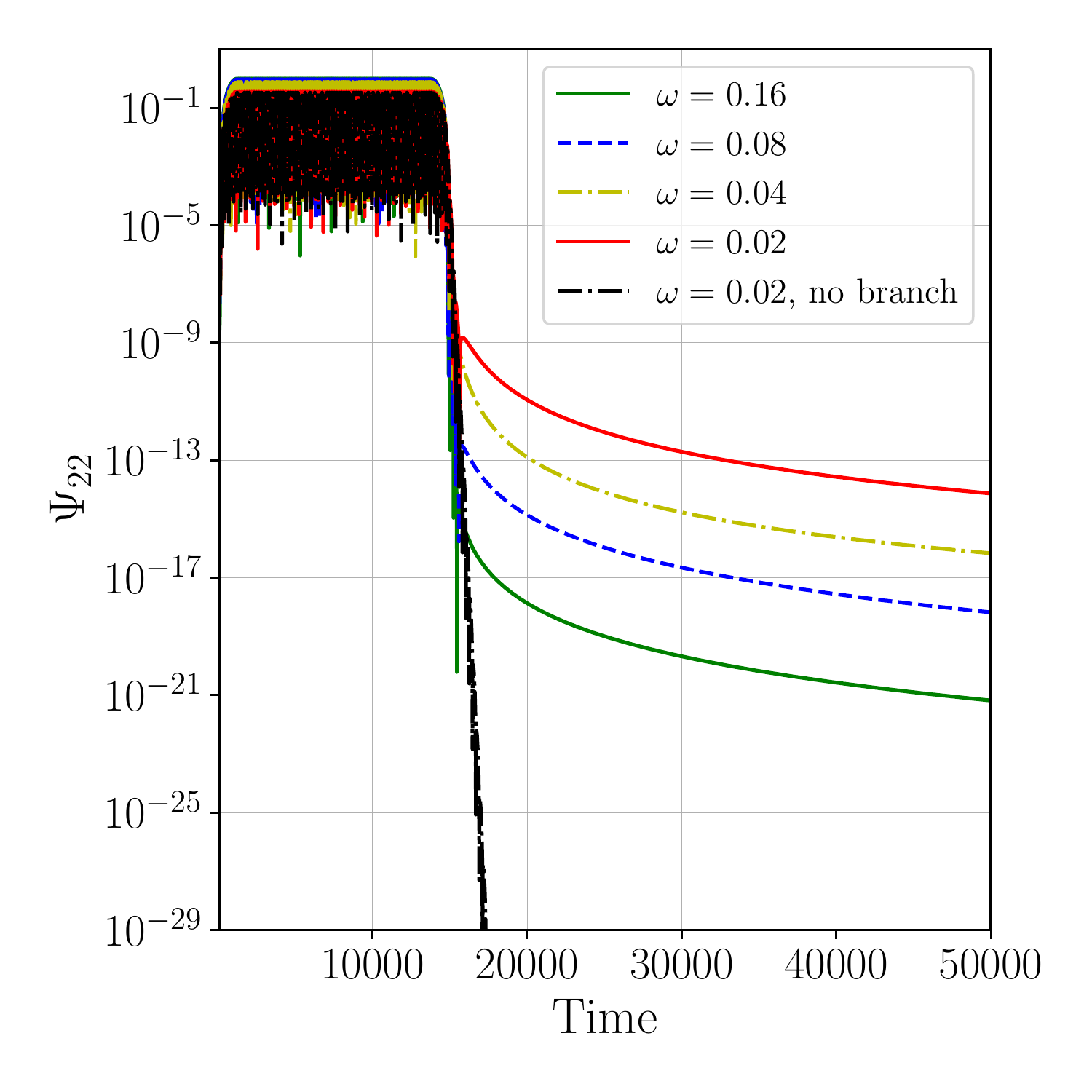}
\caption{The Zerilli function $\Psi_{22}$ at future null infinity computed 
according to Eq.~\eqref{eq:AWE}, which ``teleports'' signals recorded at $r_1=60$ to 
future null infinity. We consider a sequence of input signals at $r_1$ with increasing 
frequency $\omega$ and find that the tail generated from $r_1 \rightarrow \infty$ signal 
propagation decreases with increasing wave frequency. The tail disappears completely
when the purely real poles are discarded from the kernel's representation. These poles 
effectively approximate the kernel's branch cut along the inversion contour.}
\label{fig:AppA_kernel}
\end{figure}

\subsection{Tail excitation from circular orbits in Schwarzschild}
\label{app:simple1}

We are primarily interested in knowing how different orbital frequencies excite late-time tail behavior. To simulate non-spinning extreme mass ratio systems in a circular orbit, we numerically solve the  Regge-Wheeler-Zerilli (RWZ) equations~\cite{PhysRev1081063,PhysRevLett24737,PhysRevD71104003,Nagar:2005ea} using a high-accuracy discontinuous Galerkin solver~\cite{Field:2009kk}. In particular, we compute the $(\ell,m) = (2,2)$ Zerilli function $\Psi_{22}$ (see Eq~.1 of Ref.~\cite{Field:2009kk}) sourced by a smaller black hole orbiting a larger black hole of mass $M=1$.

We consider three kinds of circular orbits: (i) a geodesic orbit where the smaller black hole is located at $r_p =10$, (ii) a geodesic orbit where the smaller black hole is located at $r_p =30$, and (iii) a non-geodesic circular orbit where the smaller black hole is located at $r_p =10$ but whose orbital frequency is set to that of an $r_p=30$ geodesic orbit~\footnote{For circular geodesics, the orbital frequency is given by $\dot{\phi}= r_p^{-3/2}$. For the non-geodesic circular orbit, we place the particle at $r_p = 10$ but use $\dot{\phi} = 30^{-3/2}$.}. For all cases, starting at $t=5000$ we turn off the sourcing terms~\footnote{We have checked that the tail excitation is insensitive to this choice.} and monitor the amplitude $\left| \Psi_{22} \right| $ as observed at future null infinity. The far-field waveform is computed using the exact near-field-to-far-field kernel method of Ref.~\cite{Benedict:2012kw}. Our experiment's numerical parameters are exactly those reported in Sec.~4B of Ref.~\cite{Benedict:2012kw}, which in turn yields waveforms at future null infinity accurate to about $10^{-12}$ in relative error; see Table II of Ref.~\cite{Benedict:2012kw}.

\begin{figure}
\includegraphics[width=\columnwidth]{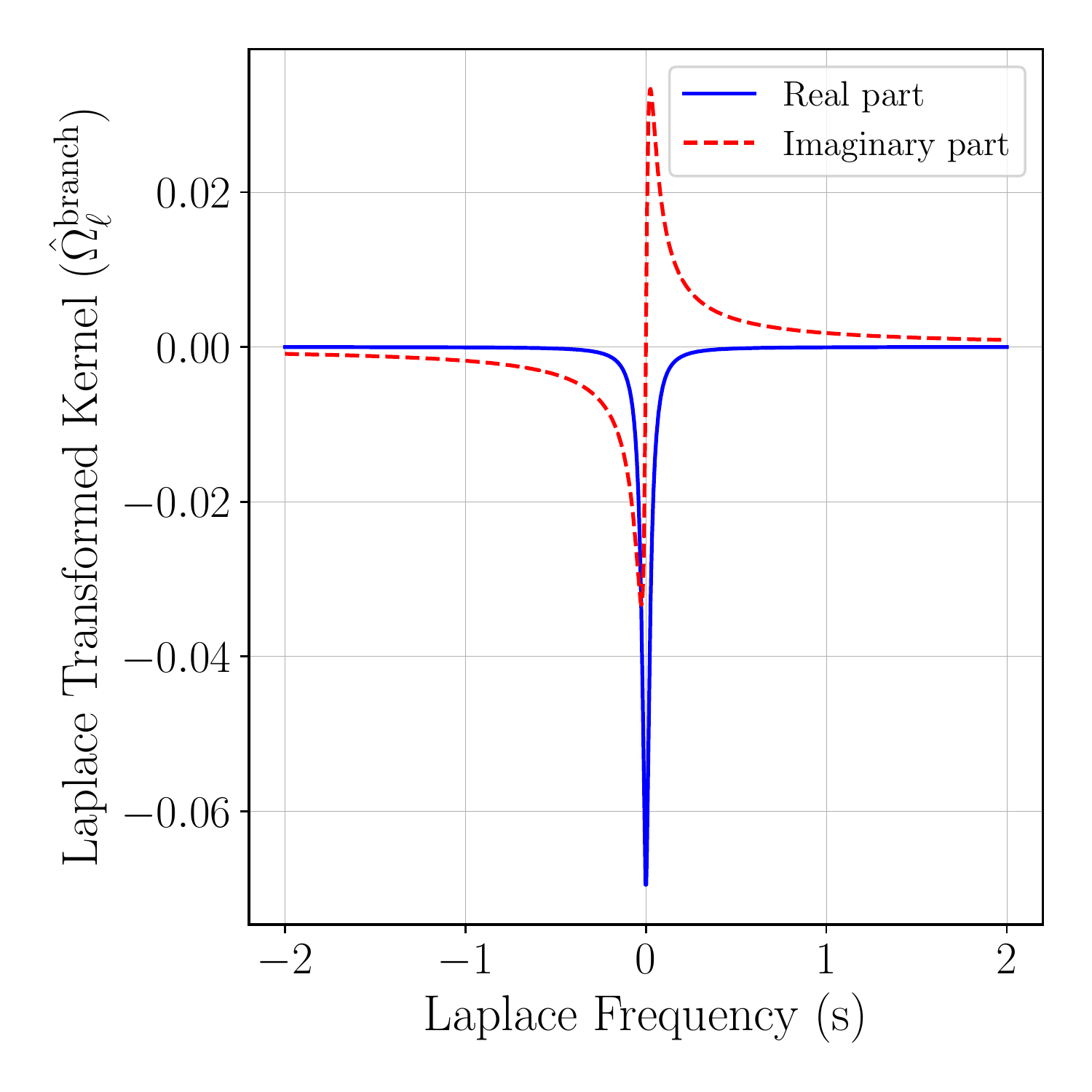}
\caption{The Laplace transformed waveform-propagation kernel, $\hat{\Omega}_{\ell}^\mathrm{branch}(s;r_1, r_2)$, evaluated along the axis of imaginary Laplace frequency (the inversion contour). This kernel enacts exact signal ``teleportation'' from $r_1=60$ to $r_2=2\times10^{15}$ according to Eq.~\eqref{eq:AWE}. Here we show only the part of the kernel that contains real poles, which we believe are responsible for approximating the kernel's branch cut along the inversion contour. The kernel's amplitude is largest in a neighborhood around $s=0$, which explains why lower frequency input signals can generate larger tails; see Fig.~\ref{fig:AppA_kernel}.}
\label{fig:AppA_kernel2}
\end{figure}

\begin{figure}[h!]
\includegraphics[width=\columnwidth]{typologies_of_plunge_geodesics.pdf}
\caption{In this figure we show the features of the different typologies of plunging geodesics considered in our work. The upper panel shows \textit{i}) a plunge geodesic with a last-apocenter-passage with $\sim 4$ whirls before plunging (orange curve), \textit{ii}) a similar plunge geodesic without whirls (dashed green curve) and \textit{iii}) a critical plunge geodesic (dotted blue curve). The first two geodesics are close to the LSO, while the latter is at the LSO.  All the plunges have $\chi=0.6$ and $e=0.9$. The bottom panel shows the time evolution of the radial coordinate of the plunges. The two plunges with a last-apocenter passage have the same radial distance at the apocenter. In the inset we highlight the features in the region where the whirls take place.}
\label{fig:trajs_geo_plot}
\end{figure}

Fig.~\ref{fig:AppA} shows the amplitude $\left| \Psi_{22} \right|$ for all three orbits after normalizing the amplitude such that they are all equal to one at $t=5000$. This normalization procedure allows us to more meaningfully compare the tail excitation between orbits. We find that the tail excitation is visually identical for orbits of the same orbital frequency regardless of their radial value $r_p$. This numerical experiment suggests that tail excitation depends on the source term's frequency, and tails are more strongly excited at lower frequencies.

\subsection{Tails generated through wave propagation}

We now consider the generation of tails as the outgoing wave propagates from some fixed radial value $r_1$ to a much larger value $r_2$. To do this, we make use of the fact that for compactly supported initial data and source terms, the solution to the Regge-Wheeler and Zerilli equation at $r=r_2$ can be written in terms of the solution at $r=r_1$ in the following form:
\begin{align} \label{eq:AWE}
& \Psi_{\ell m}(t + (r_2 - r_1),r_2)  = \nonumber \\
 & \int_0^t \Omega_{\ell}(t-\tau; r_1, r_2) \Psi_{\ell m}(\tau, r_1) d\tau + \Psi_{\ell m}(t, r_1) \,.
\end{align}
Here $\Omega_{\ell}(t; r_1, r_2)$ is a kernel (parameterized by $r_1$ and $r_2$) that can be approximated to high accuracy as a sum of damped exponentials~\cite{Benedict:2012kw}. In the Laplace frequency domain, the kernel is approximated by a sum of simple poles. These simple poles come in two flavors: (i) complex conjugate pairs and (ii) purely real. Analogous to the radiation outer boundary condition kernel~\cite{Lau:2004jn,Lau:2005ti}, we conjecture that the purely real poles approximate the effect of the kernel's branch cut (along the inversion contour) and are responsible for the generation of tails as the wave propagates from $r_1$ to $r_2$. We will use the kernel presented in Sec.~4B of Ref.~\cite{Benedict:2012kw}
for a Zerilli potential with $\ell=2$, $r_1=60$, and $r_2=2\times10^{15}$ (the location of $r_2$ is effectively future null infinity for our purposes). This kernel has 24 real poles and two complex poles (which are conjugates of one another). It is worth noting that the $\ell=2$ flatspace wave equation, for which there is no tail behavior, also has two complex poles (which are conjugates of one another) but no real poles~\cite{Field:2014cka}. This Zerilli-potential kernel was previously used to compute high-accuracy energy and angular momentum luminosity data~\cite{Benedict:2012kw}. 

Fig.~\ref{fig:AppA_kernel} shows the amplitude $\left| \Psi_{22} \right|$ of the far-field signal computed from Eq.~\eqref{eq:AWE} where the input signal is $\Psi_{\ell m}(t, 60) = \sin(\omega t)$. We taper the early part of the signal so that $\Psi_{\ell m}(0, 60) = 0$ as is required by our assumption of compactly supported initial data. We also slowly turn off the signal over the time window of 15000 to 16400. It's apparent from the figure that larger tails are generated for lower frequency waves, in line with our observations 
throughout this paper.
We also see the tails disappear entirely when we compute Eq.~\eqref{eq:AWE} after omitting the 24 real poles. This is in line with our expectation that the real poles are responsible for approximating the kernel's branch cut, while the branch cut, in turn, is responsible for generating tails as the wave propagates from $r_1$ to $r_2$. Following this insight, we can split the kernel as $\Omega_{\ell} = \Omega_{\ell}^\mathrm{branch} + \Omega_{\ell}^\mathrm{direct}$, where $\Omega_{\ell}^\mathrm{branch}$ is the part of the kernel that contains real poles and $\Omega_{\ell}^\mathrm{direct}$ contains complex poles. Further insight can be gained by Laplace transforming Eq.~\eqref{eq:AWE} to give
\begin{align} \label{eq:AWEs}
\mathrm{e}^{s(r_2-r_1)} & \hat{\Psi}_{\ell m}(s, r_2) = \nonumber \\
& \hat{\Omega}_{\ell}(s; r_1, r_2) \hat{\Psi}_{\ell m}(s, r_1) + \hat{\Psi}_{\ell m}(s, r_1) \,.
\end{align}
Eq.~\eqref{eq:AWEs} is the convolution equation~\eqref{eq:AWE} in the Laplace frequency domain. The simple algebraic 
relationship between the Laplace transformed waveform at $r_1$ and $r_2$ can be used to understand frequency-dependent tail generation. Isolating the relevant part, in Fig.~\ref{fig:AppA_kernel2}, we show $\hat{\Omega}_{\ell}^\mathrm{branch}(s;r_1, r_2)$ as a function of Laplace frequency $s$. The kernel's amplitude is largest in a neighborhood around $s=0$, which we believe explains the frequency-dependence of tail generation seen in Fig.~\ref{fig:AppA_kernel}. While this result only applies to the wave propagating from $r_1$ to $r_2$, it provides a useful intuition for the generation of tails in the late-time evolution of eccentric BBH mergers considered in Sec.~\ref{sec:tail_reason}.

\section{Typologies of Geodesic Plunges}
\label{sec:geo_trajectory}
In Figure~\ref{fig:trajs_geo_plot}, we show the three different typologies of equatorial geodesic plunges of the Kerr metric considered in our work. We use these trajectories to produce the waveforms shown in figures~\ref{fig:tail_geodesic}-\ref{fig:whirl}. In the upper panel of Fig.~\ref{fig:trajs_geo_plot} we show \textit{i}) a geodesic plunge passing through a last-apocenter-passage which has $\sim 4$ whirls before plunging (orange curve), \textit{ii}) a similar geodesic plunge with the same last-apocenter radial distance but without whirls before plunging (dashed green curve) and \textit{iii}) a critical plunge geodesics (dotted blut curve). The latter class of geodesic plunges asymptotically starts back in (coordinate) time $t$ at the unstable-circular-orbit; more details on this class of plunges can be found, for example, in Ref.~\cite{Dyson:2023fws, Faggioli2025}.  All the three typologies of plunge in Fig.~\ref{fig:trajs_geo_plot} are characterized by spin $\chi=0.6$ and eccentricity $e=0.9$. In the bottom panel of Fig.~\ref{fig:trajs_geo_plot} we show the time evolution of the radial coordinate of the three plunge geodesics: we stress that the two geodesics with a last-apocenter-passage have an almost identical evolution up to reaching the radius of the unstable circular orbit with the same angular momentum, where they differ by the number of whirls performed before continuing their plunge.

We construct these geodesics using the generic analytic solutions for plunge geodesics obtained in~\cite{Dyson:2023fws} as implemented in \texttt{KerrGeoPy}~\cite{KerrGeoPy}. The equatorial plunge solutions are parameterized by their energy and angular momentum. We therefore need two conditions to fix a single plunge. The first condition is that we fix the apocenter distance (\gu{51.9}M in the plotted example). For the second we fix the difference between the energy of the geodesic and the energy of the unstable circular orbit with the same angular momentum. The number of whirls scales approximately with the logarithm of this difference~\cite{Glampedakis:2002ya, Pretorius:2007jn}. By making the difference arbitrarily small we can, in principle, complete arbitrarily many whirls. In practice, we find that the analytic solutions with more than $\sim 4$ whirls require more than machine precision to accurately evaluate. Hence, we chose a solution with approximately 4 whirls. We numerically root find these two conditions to find the corresponding energy and angular momentum.

\section{Influence of last apocenter passage}
\label{sec:influence_of_last_apocenter}
\begin{figure}
\includegraphics[width=\columnwidth]{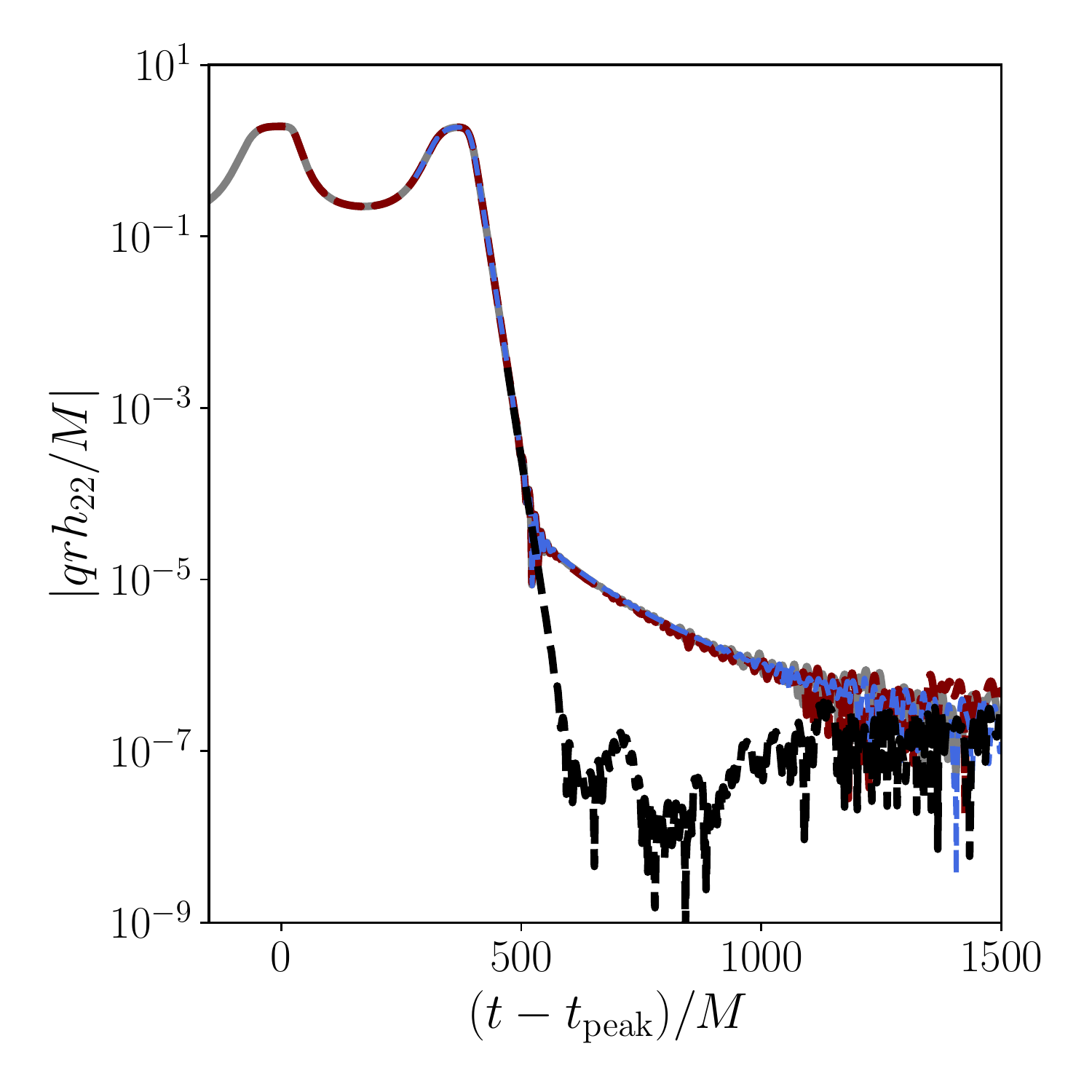}
\caption{We show the $(2,2)$ amplitude generated using RR-force driven trajectories. Each trajectory has the same eccentricity $e=0.9$ and spin $a=0.6$ and is calculated using different starting times $t_{\rm start}$ for the evolution. We observe that when the trajectory does not perform a last-apocenter-passage, the tail is not present in the (2,2) ringdown amplitude (black dashed line). When the tail exists, we do not observe a relevant dependence of the tail features with respect to the starting time of the trajectory evolution.}
\label{fig:different_starting_time_tail}
\end{figure}

In this section we perform an experiment that highlights how a last-apocenter pass of the trajectory is necessary to recover a late-time tail in the waveform. We consider a set of trajectories where evolution is driven by the RR-force introduced in Sec.~\ref{sec:BHPT_simulation} and that is characterized by the same eccentricity $e=0.9$ and spin $a = 0.6$. For every trajectory, we start the evolution from different starting times $t_{\rm start}$ before the merger, so that at least one trajectory does not have a last-apocenter-pass. In Fig.~\ref{fig:different_starting_time_tail} we show the $(2,2)$ amplitude of the waveforms produced with these trajectories. We observe that whenever the trajectory does not perform a last-apocenter-passage, the amplitude of the relative waveform does not exhibit late-time tail. On the other hand, when the tail exists we find that its features do not depend on the starting time of the trajectory evolution. 

\bibliography{References}

\end{document}